\documentclass[pre,aps,amsfonts,amssymb,amsmath,graphicx,showpacs,epsfig,amscd]{revtex4}
\usepackage{graphicx}
\usepackage{amsmath}
\usepackage{amsfonts}
\usepackage{amssymb}
\def\bi{\begin{itemize}}

\def\ei{\end{itemize}}
\def\be{\begin{equation}}
\def\ee{\end{equation}}
\def\bea{\begin{eqnarray}}
\def\eea{\end{eqnarray}}

\DeclareTextSymbol{\degre}{OT1}{23}

\begin{document}

\author{G. Ovarlez$^{1,2}$\footnote{author to whom correspondence should be addressed: ovarlez@lcpc.fr}, E. Cl\'{e}ment$^2$}
\affiliation{$^1$ Laboratoire des Mat\'eriaux et Structures du
G\'enie Civil - UMR 113 (LCPC-ENPC-CNRS)\\ Institut Navier - 2,
all\'ee Kepler - 77420 Champs sur Marne - France\\ $^2$ Physique
et M\'{e}canique des Milieux H\'{e}t\'{e}rog\`{e}nes (UMR 7636
CNRS-ESPCI) and Universit\'{e} Pierre et Marie Curie \\ ESPCI -
10, rue Vauquelin - 75231, Paris Cedex 5 - France }
\title{Elastic medium confined in a column versus the Janssen experiment.}
\date{\today}
\begin{abstract}
We compute the stresses in an elastic medium confined in a vertical column,
when the material is at the Coulomb threshold everywhere at the walls.
Simulations are performed in 2 dimensions using a spring lattice, and in 3
dimensions, using Finite Element Method. The results are compared to the
Janssen model and to experimental results for a granular material. The
necessity to consider elastic anisotropy to render qualitatively the
experimental findings is discussed.
\end{abstract}
\maketitle

\section{Introduction}

The mechanical status of granular matter is presently one of the
most open and debated issues \cite{PDM}. This state of matter
exhibits many unusual mechanical and rheological properties such
as stress induced organization at the microscopic \cite{Oda} or at
the mesoscopic \cite{Radjai} level which may yield macroscopic
effects such as stress induced anisotropy \cite{Geng03,Attman04a}.
This issue sets fundamental questions relevant to the
understanding of many other systems exhibiting jamming such as
dense colloids or more generally soft glassy materials
\cite{Nagel,Cates}. For practical applications, the quasistatic
rheology of granular assemblies is described using a
phenomenological approach, based on an elasto-plastic modelling of
stress-strain relations \cite{Wood}. So far, there is no consensus
on how to express correctly the macroscopic constitutive relations
solely out of microscopic considerations and under various
boundary conditions or loading histories. This very basic issue
was illustrated in a recent debate on how to understand the stress
distribution below a sand pile and especially how to account for
the dependence on preparation protocols \cite{Vaneltroudutas}. A
new mechanical approach was proposed based on the concept of
''fragile matter'' \cite{Cates} and force chains propagation
modelling \cite{OSL}. But recent experiments have dismissed this
approach and evidenced results more consistent with the
traditional framework of general elasticity \cite{Guiguir}.

In this paper we focus on the predictions for stresses
measurements at the bottom for an elastic material confined in a
rigid cylinder. When the column is filled with granular material
it corresponds to the classical Janssen's problem \cite{Janssen}.
Recently, this issue has received a lot of attention either
experimentally
\cite{Vaneljanssen,OvarlezRheo01,OvarlezRheo03,OvarlezSurp03,Kolb99,bertho02,Zenith03}
or numerically \cite{landry04,Radjai04}, the confined material
being either pushed or pulled down. Surprisingly, so far to our
knowledge, there are very few systematic comparison or even direct
relation with the outcome of standard elasticity in the same
situation of confinement. Note that experimentally, it was found
that essentially elastic materials like gels, may display a
Janssen stress saturation effects that could well predict the
onset of self-collapsing under gravity \cite{Allain}. In a recent
paper \cite{Evesque}, Evesque and de Gennes proposed a model for
the slow filling of an elastic medium modelling a granular
packing. As the material is poured in the column, displacements of
the material at the bottom are induced by the weight of the new
material added so that it mobilizes friction forces. Within the
assumption of a minimal anchorage length, it leads to partial and
inhomogeneous mobilization of friction at the walls: friction is
fully mobilized at the bottom, and partially in the upper part of
the column. A Janssen like pressure profile can be derived in the
case where the saturation length $\lambda$ is high compared to the
column radius $R$; this last condition is actually restrictive and
inappropriate for usual experimental cases
\cite{Vaneljanssen,OvarlezSurp03}.

We propose to study in detail the elastic predictions and to compare them to
the Janssen model and to experimental results for a granular material in
\textit{the same situation} i.e. when the material is at the Coulomb threshold
\textit{everywhere} at the walls. The main comparison features with the
experimental results have already been presented in \cite{OvarlezSurp03};
here, we detail much more the elastic predictions.

Two kinds of situations are considered. First, the mass at the bottom of the
column is measured as a function of the material filling mass. Second, similar
measurements are produced with an overweight on the top of the material. We
recently performed the corresponding experiments for a granular material
\cite{OvarlezSurp03} and it was shown that one could obtain quite reproducible
data provided a good control of the packing fraction homogeneity and a
polarization of all the friction forces at the walls in the upwards direction.
Our measurements confirmed, for the first time, the general validity of
Janssen's saturation curve. They also evidenced an overshoot effect of
spectacular amplitude induced by a top mass equal to the saturation mass.
These experimental results are actually strong tests for any theory of
granular matter.

\section{Experimental results and Janssen model}

\begin{figure}[ptb]
\includegraphics[width=6cm,angle=-90]{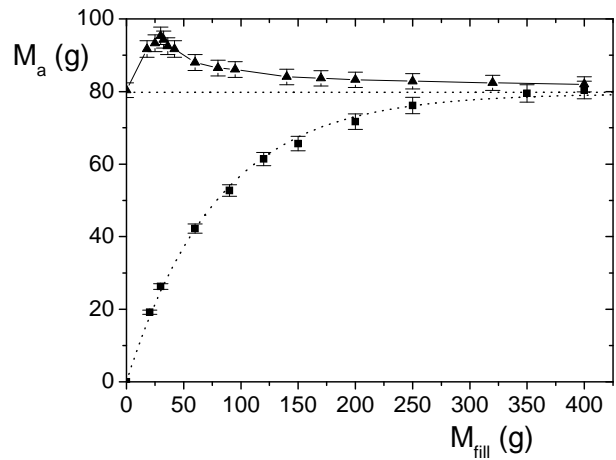}\caption{Apparent mass
$M_{a}$ vs. filling mass $M_{fill}$ for a loose packing
($\bar\nu=59\%$) of slightly polydisperse 1.5 mm glass beads in a
38 mm diameter steel column of friction coefficient $\mu_{s}=0.25$
without an overweight (squares), and with a 80.5 g overweight
(triangles, line) on top of the granular material. The results of
both experiments are compared to the Janssen model prediction
(dotted lines).} \label{fig1}
\end{figure}

We first summarize our recent experimental results
\cite{OvarlezSurp03} and compare them to the Janssen model
predictions. In \cite{OvarlezSurp03}, a slightly polydisperse
assembly of 1.5 mm glass beads was poured at controlled packing
fraction in steel columns of various friction coefficient, and of
diameter varying between 38 mm and 80 mm. The aim of the
experimental procedure is to achieve the Coulomb threshold
everywhere at the walls when the apparent mass $M_{a}$ is
measured; see \cite{OvarlezSurp03} for details on the procedure.
The apparent mass $M_{a}$ measured at the bottom of the column was
plotted as a function of the filling mass $M_{fill}$ of the
material. The typical results obtained when $M_{fill}$ is varied
are shown on Fig. \ref{fig1}. The apparent mass $M_{a}$ saturates
exponentially with $M_{fill}$. When an overweight equal to the
saturation mass $M_{sat}$ is added on top of the granular
material, $M_{a}$ increases with $M_{fill}$, up to a maximum
$M_{max}$, which is about $20\%$ higher than $M_{sat}$, then
decreases slowly towards the saturation mass $M_{sat}$.

The simple model which captures the physics of this saturation phenomenon was
provided in 1895 by Janssen \cite{Janssen}. This model is based on the
equilibrium of a granular slice taken at the onset of sliding everywhere at
the walls; we attempted to realize as best as possible this last condition in
our experiment \cite{OvarlezSurp03}. In cylindrical coordinates with origin at
the top surface and the cylinder axis being the $z$ axis, the relation, at the
slipping onset, between the shear stress $\sigma_{rz}$ and the horizontal
stress $\sigma_{rr}$ at the walls is
\begin{equation}
\sigma_{rz}(r\!\!=\!\!R,z)=\mu_{s}\sigma_{rr}(r\!\!=\!\!R,z)
\end{equation}
where $\mu_{s}$ is the Coulomb static friction coefficient between the grains
and the walls. It results in a relation between the filling mass $M_{fill}$
and the apparent mass at the bottom $M_{a}$ of the form:
\begin{equation}
M_{a}=M_{sat}(1-\exp(-\frac{M_{fill}}{M_{sat}})) \label{janssen}
\end{equation}
with
\begin{equation}
M_{sat}=\frac{\rho\pi R^{3}}{2K\mu_{s}}
\end{equation}
where $\rho$ is the mass density of the granular material, and $K$ is the
Janssen parameter rendering the average horizontal redirection of vertical
stresses:
\begin{align}
\sigma_{rr}=K\sigma_{zz} \label{redirection}
\end{align}
From a mechanical point of view, a major simplification of this model comes
from the assumption that the redirection parameter $K$ would stay constant
along the vertical direction. But on the other hand, it provides a clear and
simple physical explanation for the existence of an effective screening length
$\lambda$ $=R/2K\mu_{s}$ above which the mass weighted at the bottom saturates.

In \cite{OvarlezSurp03}, several saturation profiles were measured
for various packing fractions, columns sizes and friction
coefficients between the grains and the walls. When the apparent
mass rescaled by the saturation mass is plotted as a function of
the filling mass also rescaled by the saturation mass, we obtain a
\textit{universal rescaling} of all data on a curve which is
precisely the one predicted by Janssen (Fig. \ref{fig1}):
$M_{a}/M_{sat}=f(M_{fill}/M_{sat})$, with $f(x)=1-\exp(-x)$. The
rescaling with radius $R$ and friction coefficient $\mu_{s}$ was
also checked, and good agreement with the Janssen model rescaling
was found. The Janssen constant $K$ was found to depend on packing
fraction $\bar\nu$ and an effective relation was derived: $\Delta
K/K\simeq5\Delta\bar\nu/\bar\nu$. On the other hand, when a top
mass equal to the saturation mass is added on the top of the
granular material, the apparent mass $M_{a}$ displays a maximum
$M_{max}$ $20\%$ higher than $M_{sat}$, whereas the Janssen model
predicts $M_{a}=M_{sat}$ whatever the filling mass $M_{fill}$ is.
Therefore, this overshoot goes beyond the possibilities of
Janssen's model which seems adapted to a unique configuration.

In the next section, we study in detail the predictions of isotropic
homogeneous elasticity.

\section{Elasticity: theory and simulation methods}

\subsection{Theory}

\label{elastheory} We first recall the general framework of homogeneous
isotropic elasticity, and then predict the behavior of an elastic material
confined in a column.

The elastic theory gives, in the limit of small deformations, a linear
relation between the stress tensor components $\sigma_{ij}$ and the strain
tensor components $\epsilon_{ij}$. For an isotropic elastic material, we get
\begin{equation}
E\epsilon_{ij}=(1+\nu_{p})\sigma_{ij}-\nu_{p}\delta_{ij}\sigma_{kk}
\label{stressstrain}
\end{equation}
where E is the Young modulus, and $\nu_{p}$ the Poisson ratio which takes its
value between $-1$ and $1/2$ in 3D, and between -1 and 1 in 2D.

\begin{figure}[ptb]
\includegraphics[width=8cm]{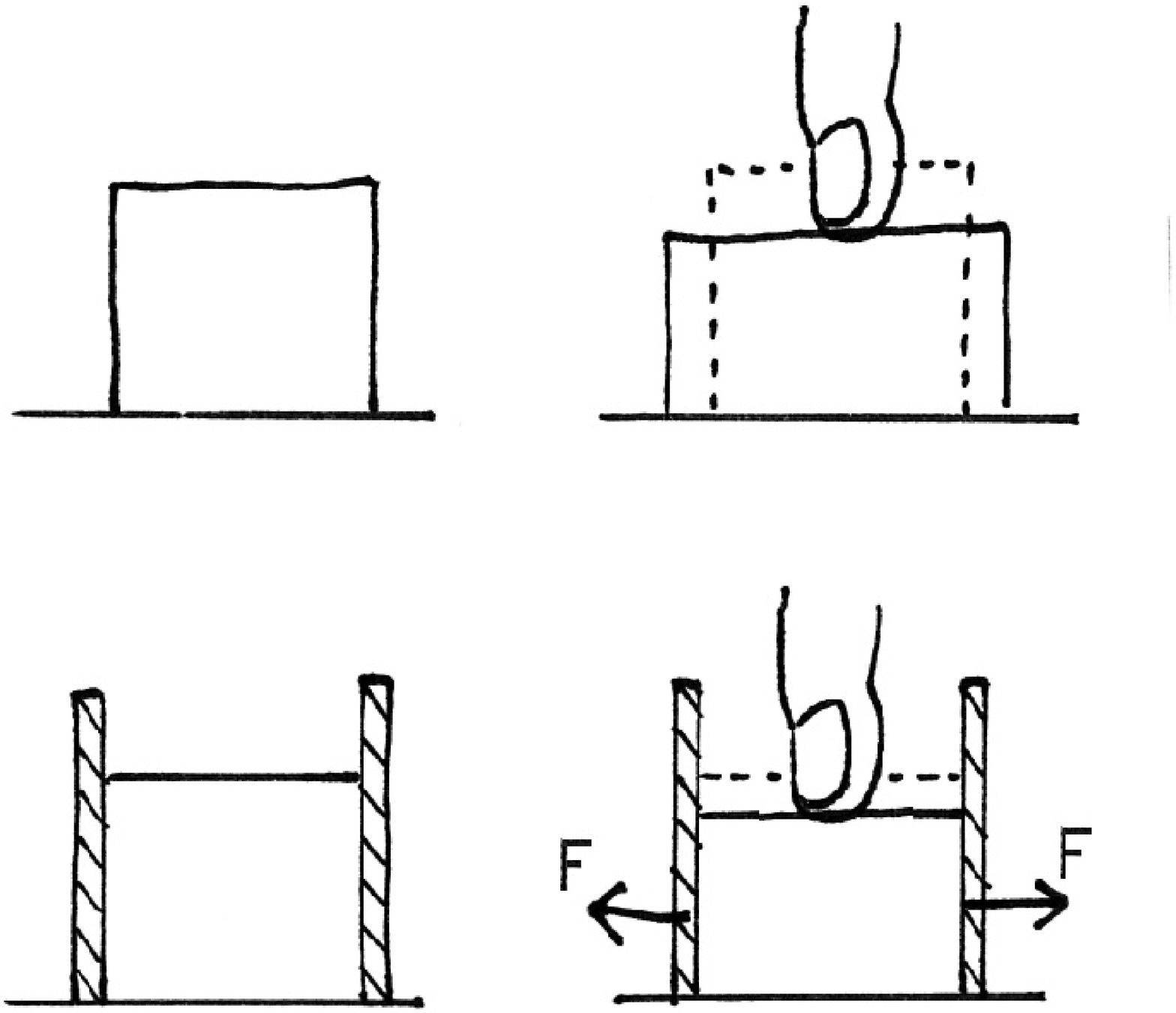}
\caption{Sketch of uniaxial
compression of a free elastic material (top) and of a confined
elastic material (bottom).} \label{Fig2}
\end{figure}

In a uniaxial homogeneous compression experiment (Fig.
\ref{Fig2}), where $\sigma_{zz}=-p$ is imposed everywhere, the
other stress tensor components being null, we get
\begin{equation}
\epsilon_{zz}=-p/E
\end{equation}
everywhere and
\begin{equation}
\epsilon_{xx}=\epsilon_{yy}=-\nu_{p} \epsilon_{zz}
\end{equation}
which signifies that the material expands in the transverse
direction.\newline The Young modulus $E$ is thus characteristic of the
material's stiffness; a cylinder of length $l$ and section $S$ has stiffness
$k=ES/l$ in the axial direction. The Poisson ratio is linked to the material
compressibility: the volume variation is: $\delta V/V=-(1-2\nu_{p})p/E$.
Therefore, an incompressible material has Poisson ratio $1/2$ in 3D (1 in 2D).

If we now confine an elastic medium of Young modulus $E$ and Poisson ratio
$\nu_{p}$, in a rigid cylinder of radius $R$, no more radial displacement at
the walls is allowed: $u_{r}(r\!=\!R)=0$. Stresses and displacements can
actually be calculated in the \textit{limit of high depths $z$} under the
assumption that they then should be independent of $z$. The boundary
conditions we impose are the Coulomb condition everywhere at the walls
\begin{equation}
\sigma_{rz}(r\!\!=\!\!R)=\mu_{s}\sigma_{rr}(r\!\!=\!\!R)
\end{equation}
and infinitely rigid walls i.e. $u_{r}(r\!\!=\!\!R)=0$. The stress tensor
components are then
\begin{align}
\sigma_{rz}(r,z)  &  =-\frac{1}{2}\rho g r\\
\sigma_{rr}=\sigma_{\theta\theta}  &  =\frac{\nu_{p}}{1-\nu_{p}}\;\sigma
_{zz}\\
\sigma_{zz}^{sat}(r,z)  &  =-\frac{(1-\nu_{p})\rho
gR}{2\nu_{p}\mu_{s}} \label{stressasympt}
\end{align}
And the asymptotic displacements are
\begin{align}
u_{z}(r,z)  &  =-\frac{1\!+\!\nu_{p}}{2E}\rho gr^{2}-\frac{1\!-\!\nu
_{p}\!-\!2\nu_{p}^{2}}{2\mu_{s}\nu_{p}E}\rho gRz+u_{0}\label{depasympt}\\
u_{r}(r,z)  &  =u_{\theta}(r,z)=0
\end{align}
This can be checked by injecting this solution in the stress-strain relation
(\ref{stressstrain}) and internal equilibrium relation
\begin{equation}
\partial_{i}\sigma_{ij}=-\rho g_{j}
\end{equation}

Thus, we obtain a Janssen's like redirection phenomenon due to a
Poisson's ratio effect with a local Janssen's parameter
$K_{el}=\sigma_{rr}(r,z)/\sigma_{zz}(r,z)$, being for large
depths:
\begin{equation}
K_{el}=\frac{\nu_{p}}{1-\nu_{p}}
\end{equation}
At 2D, we obtain the same saturation and stress redirection phenomena with
$K_{el}=\nu_{p}$.

For a free elastic medium, the Poisson ratio effect is a
transverse dilatation; for a confined elastic material, the
Poisson ratio effect is a transverse redirection of stresses (Fig.
\ref{Fig2}).

In the following, $K_{el}$ will design the elastic stress redirection
constant, whereas $K$ is devoted to design the stress redirection constant in
the Janssen framework.

In this section, we obtained the asymptotic values of stresses and
displacements. For the vertical stress, the limit is similar to the Janssen
asymptotic vertical stress if one identifies $K$ an $K_{el}$. But we cannot
say anything from this calculation about the whole pressure profile. We thus
need to perform numerical computation. In the next section, we present the
numerical methods we employed to simulate Janssen experiments for an elastic column.

\subsection{Numerical methods}

Two different methods were employed: we first computed the stresses in 2D with
a spring lattice. We also computed the stresses in 3D, using Finite Element
Method thanks to CAST3M \cite{Castem}.

\subsubsection{2D: spring lattice}

The 2D computations are performed in order to provide elastic predictions for
a direct comparison with 2D simulations of granular materials.

In order to avoid any confusion, let us recall that we are not trying here to
give a microscopic model for a granular material, but we use a discrete system
behaving like an effective elastic medium in the continuous limit in order to
perform simple numerical simulations.

However, this system can describe the most simple of granular materials: a 2D
hexagonal piling of frictionless disks (with non-hertzian contacts so that
contact elasticity is linear). Note also that systems of springs have been
recently studied by Goldenberg and Goldhirsch \cite{Elastochaines1}; they
showed that large forces inhomogeneities (i.e. like forces chains) can be
found at the discrete scale in these systems, which are elastic in the
continuous limit.

\begin{figure}[ptb]
\includegraphics[width=7cm]{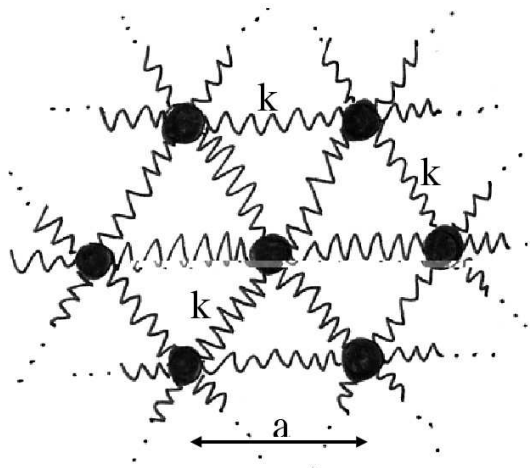}\caption{Sketch of the discrete 2D
elastic medium.} \label{fig3}
\end{figure}

In order to simulate a 2D elastic medium, we put point masses $m$
on an hexagonal lattice of link size $a$ (Fig. \ref{fig3}). Every
particle is linked to her 6 neighbors with identical springs of
stiffness $k$ and length $a$ at rest. Therefore, the potential
interaction energy between particles placed at $0$ and $x_{i}$
(such that $x_{i}x_{i}=a^{2}$) at rest, submitted to infinitesimal
displacements $u_{i}$ and $v_{i}$, is:
\begin{equation}
E_{p}=\frac{1}{2}k
\Bigl(\sqrt{(x_{i}+v_{i}-u_{i})(x_{i}+v_{i}-u_{i})} -a\Bigr)^{2}
\end{equation}

\begin{figure}[ptb]
\includegraphics[width=8cm]{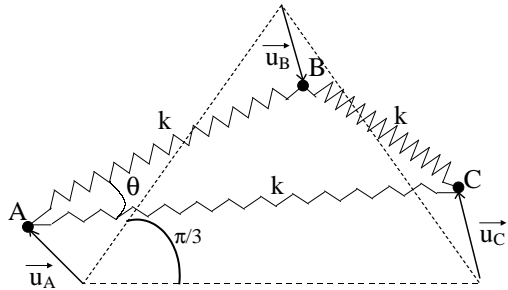}\caption{Displacements of the points
of an elementary cell of the 2D elastic medium.} \label{fig4}
\end{figure}

By varying the stiffness $k$, we can only vary the young modulus
$E$ of the effective elastic medium. As we also need to vary the
Poisson ratio $\nu$, an elastic torsion potential between neighbor
springs separated by angle $\theta$ (Fig. \ref{fig4}) is added:
\begin{equation}
E_{p}=\frac{1}{2}k_{b}\cos^{2}(\theta-\frac{\pi}{3})
\end{equation}
This potential tries to maintain an angle $\pi/3$ between 2
neighbors springs if $k_{b}<0$. Thus the limit
$k_{b}\rightarrow\-infty$ corresponds to a contractive medium of
Poisson ratio -1, whereas the limit $k_{b}\rightarrow\infty$
corresponds to a incompressible medium.

An elementary area $A=a^{2}\sqrt{3}/2$ can be associated to each particle. The
surface energy $\omega$ then reads:
\begin{equation}
\omega=\frac{2}{\sqrt{3}a^{2}}\biggl(\frac{1}{2}\sum E_{p}
(springs)+\sum E_{p}(angles)\biggr)
\end{equation}

If we consider the continuum limit of this system, the displacement of a
particle located at $x_{i}$ is:
\begin{equation}
u_{i}=(\partial_{j}u_{i}) x_{j}
\end{equation}
The surface potential energy can be easily computed at second order in
$\epsilon_{ij} =(1/2)(\partial_{i}u_{j}+\partial_{j}u_{i})$, and one obtains:
\begin{equation}
\omega=
\frac{\sqrt{3}}{2}\biggl(\frac{3}{4}(k+2k_{b})(\epsilon_{xx}^{2}+\epsilon_{yy}^{2})+\frac{1}{2}(k-6k_{b})\epsilon_{xx}\epsilon
_{yy}+(k+6k_{b})\epsilon_{xy}^{2}\biggr)\end{equation}

For an elastic medium, stress-strain linearity reads
\begin{equation}
\sigma_{ij}=C_{ijkl}\epsilon_{kl}\end{equation}

There are \textit{a priori} 9 independent parameters in $C_{ijkl}$
(but only 2 if it is an isotropic elastic medium). The surface
energy then reads
\begin{equation}
\omega=\frac{1}{2}\sigma_{ij} \epsilon_{ij}=\frac{1}{2}
\epsilon_{ij} C_{ijkl}\epsilon_{kl} \label{omega}
\end{equation}

If we identify this energy to the one we obtained for the spring lattice, we
see that our system, in the continuous limit, is an isotropic elastic medium
of Young modulus E and Poisson ratio $\nu_{p}$:
\begin{align}
\nu_{p}  &  =\frac{1}{3}\frac{k-6k_{b}}{k+2k_{b}}\\
E  &  =\frac{2\sqrt{3}}{3}k\frac{k+6k_{b}}{k+2k_{b}}
\end{align}

In order to solve the equilibrium problem of the lattice, we need
to express the internal forces. They can be deduced from the
surface energy (\ref{omega}): to the usual elastic forces due to
compression (or decompression) of springs, add forces which tends
to drive the triangles angles to their equilibrium value (if
$k_{b}<0$) or out of the $\pi/3$ value (if $k_{b}>0$). The force
on A, due to the out of equilibrium angles of triangle ABC (Fig.
\ref{fig4}) reads:
\begin{equation}
F_{A_{i}}=-\frac{3}{2}k_{b} (u_{A_{i}}+R_{{-\pi/3}_{ij}}u_{B_{j}}+R_{{\pi
/3}_{ij}}u_{C_{j}})
\end{equation}
where $R_{{\theta}_{ij}}$ is the rotation matrix of angle $\theta$.

For the numerical computation, we impose the balance of forces (gravity,
elastic forces, torsion forces) everywhere. At the bottom, we impose a null
vertical displacement (rigid bottom), and either a perfectly stick (i.e.
nullity of horizontal displacement) or perfectly slip (i.e. nullity of
horizontal projection of forces) bottom. At the walls, we impose a null
horizontal displacement (rigid wall), and we impose the Coulomb condition: the
forces projected vertically are proportional to the forces projected
horizontally with proportionality factor $\mu_{s}$. On top of the material,
the forces projected horizontally are null (perfectly slip overweight), and
the overweight is simulated by imposing the same vertical displacement for
each particle on top (stiff overweight), which is close to the experimental
situation. We thus obtain a linear system on the point displacements. We can
vary stiffness $k$ and $k_{b}$ in order to simulate elastic mediums of
different Young modulus and Poisson ratio. Note that for varying $\nu_{p}$
between 0 and 1, we need to vary $k_{b}$ between $-k/6$ and $k/6$. We also
vary the friction at the walls $\mu_{s}$.

\subsubsection{3D: FEM}

In order to get the whole stress saturation curve, finite element numerical
simulations \cite{Castem} were performed. The column is modelled as an
isotropic elastic medium. We vary the friction at the walls $\mu_{s}$, the
Young modulus $E$ and the Poisson ratio $\nu_{p}$. We imposed a rigid (nullity
of vertical displacements), either perfectly stick (nullity of horizontal
displacements) or perfectly slip (nullity of horizontal stresses) bottom. We
found no appreciable difference between these two previous cases. The
condition $\sigma_{rz}=\mu_{s}\sigma_{rr}$ is imposed everywhere at the walls.
The cylinder is modelled as a steel elastic medium. We verified that in all
the simulations performed, there is no traction in the elastic medium, so that
this can actually be a model for a granular material.

For the simulations performed without overweight, the top surface is set free
(no stress); the overweight is modelled as a perfectly slip (no horizontal
stress) brass elastic medium.

In order to impose the Coulomb condition at the walls, we first set $F_{z}=0$
at each point of the mesh at the walls; we then obtain in these conditions the
value $F_{r}$ exerted by the elastic medium on the walls. We then iterate in
order to obtain $F_{z}=\mu_{s} F_{r}$ where $\mu_{s}$ is the friction
coefficient at the walls: the vertical force imposed at step (i+1) is:
\begin{equation}
F_{z}(i+1)=(1-\epsilon)\times F_{z}(i) + \epsilon\times\mu_{s} F_{r}(i)
\end{equation}
We choose $\epsilon=0.2$. This procedure ensures convergence towards the
Coulomb condition: if at step (i) the Coulomb condition $F_{z}(i)=\mu_{s}
F_{r}(i)$ is fulfilled, then at step (i+1): $F_{z}(i+1)=(1-\epsilon) F_{z}(i)
+ \epsilon\mu_{s} F_{r}(i)=(1-\epsilon)\mu_{s} F_{r}(i) + \epsilon\mu_{s}
F_{r}(i)=\mu_{s} F_{r}(i)=F_{z}(i)$. This boundary condition is the same as
the one at step (i): this yields $F_{r}(i+1)=F_{r}(i)$, and thus
$F_{z}(i+1)=\mu_{s} F_{r}(i+1)$.

\subsubsection{Remarks}

Note that in these simulations we imposed the Coulomb condition everywhere at
the walls. This allows comparison with the Janssen model in the same
situation. But, regarding the experimental results \cite{OvarlezSurp03}, i)
nothing really insures that our piling preparation is strictly isotropic and
ii) in spite of the careful procedure, we are never absolutely sure that all
the friction forces at the wall are actually mobilized upwards. Moreover, the
modelling of the contacts may seem rudimentary, as elasticity of contact
should be included.

In these simulations, imposing dynamical friction at the walls or the static
Coulomb threshold is actually the same: the material obeys the same
equilibrium equations (if we consider a steady sliding at the walls), and the
same condition at the walls, with just a change in the name (and the
experimental value) of the friction coefficient.

\section{Simulation results}

In this section, we present the results obtained from numerical computations
at 2D and 3D.

In the following, we vary mainly the friction coefficient, and the Poisson
ratio. The 2D simulations are in arbitrary units. The 3D simulations were all
performed, if no contrary mention, for an isotropic elastic medium of mass
density $\rho=1.6\ \mbox{g\,cm}^{-3}$ (which corresponds to an assembly of
glass beads of packing fraction $\bar\nu=64\%$), of Young modulus 100 MPa, in
a steel cylinder of radius $R=4$ cm, Poisson ratio 0.3, Young modulus 210 GPa,
and thickness 3 mm. These data, which correspond to the display used in
\cite{OvarlezSurp03}, will not be specified anymore in the following.
Simulations will also be performed in order to study the influence of the
variation of the cylinder radius and the Young modulus of the elastic medium.

\subsection{Simulations without overweight}\label{Nooverweight}

\begin{figure}[ptb]
\includegraphics[width=8cm]{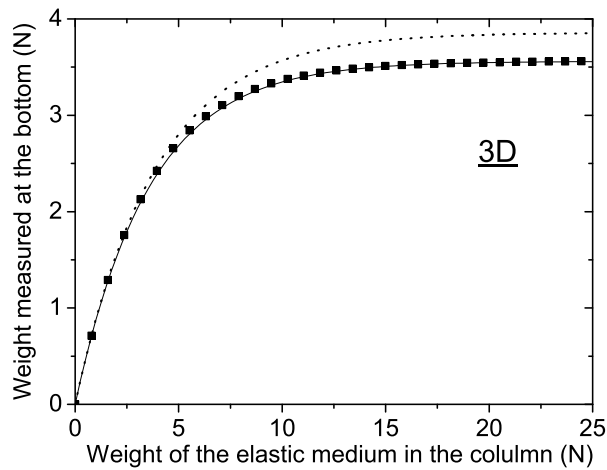}\caption{3D simulation of a
Janssen experiment (squares) for an elastic material of Poisson
ratio $\nu _{p}=0.45$; the friction at the walls is $\mu_{s}=0.5$.
The data are fitted by a Janssen curve of coefficient $K=0.89$
(line). The Janssen curve for $K$ corresponding to the elastic
stress redirection constant $K_{el}=\nu _{p}/(1-\nu_{p})=0.82$ is
also displayed (dotted line).} \label{fig5}
\end{figure}

We first perform simulations similar to the original Janssen
experiment: we plot the weight at the bottom as a function of the
weight of elastic material in the column (Fig. \ref{fig5}).

We see on Fig. \ref{fig5}, for friction coefficient $\mu_{s}=0.5$
and Poisson ratio $\nu_{p}=0.45$, that the apparent mass $M_{a}$
saturates exponentially with the filling mass. The data are
perfectly fitted by the Janssen model for these parameters (Fig.
\ref{fig5}), but the Janssen coefficient $K$ extracted from the
fit is $9\%$ higher than the elastic stress redirection constant
$K_{el}=~\nu_{p}/(1-\nu_{p})=0.82$. This is \textit{a priori}
unexpected as the elastic saturation pressure and the Janssen one
should be identical with $K=K_{el}$.

\subsubsection{Effect of the bottom}

\begin{figure}[ptb]
\includegraphics[width=8cm]{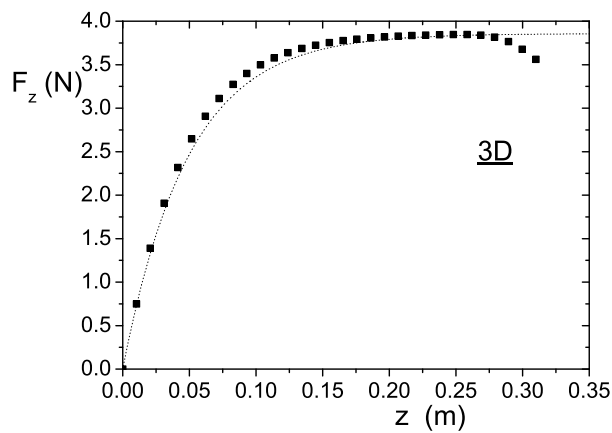}\caption{Mean pressure profile
in a simulated elastic material of height 31 cm and Poisson ratio
$\nu _{p}=0.45$ (squares); the friction at the walls is
$\mu_{s}=0.5$. Depth $z=0$ cm corresponds to the top of the
column. Depth $z=31$ cm corresponds to the bottom. We display the
integral of vertical stresses $F_{z}$ at height $z$. The data are
compared to a Janssen curve of coefficient $K=K_{el}=0.82$
corresponding to the elastic stress redirection constant (dotted
line).} \label{fig6}
\end{figure}

In order to understand this feature, we study the whole mean
vertical pressure profile in the same column (Fig. \ref{fig6}).

Regarding mean vertical pressure, we see that the asymptotic value
is now the expected theoretical asymptotic value, and the Janssen
curve with $K=K_{el}$ gives a good though not perfect fit of the
profile. The pressure saturates at high depths but decreases
suddenly near the bottom; this is actually the value on the bottom
we measure in a Janssen experiment. This feature explains why the
saturation mass is lower than the expected one on Fig. \ref{fig5}.
The reason for this change of pressure near the bottom is that the
asymptotic vertical displacement is parabolic whereas the bottom
is rigid and imposes a flat displacement.

\begin{figure}[ptb]
\includegraphics[width=10cm]{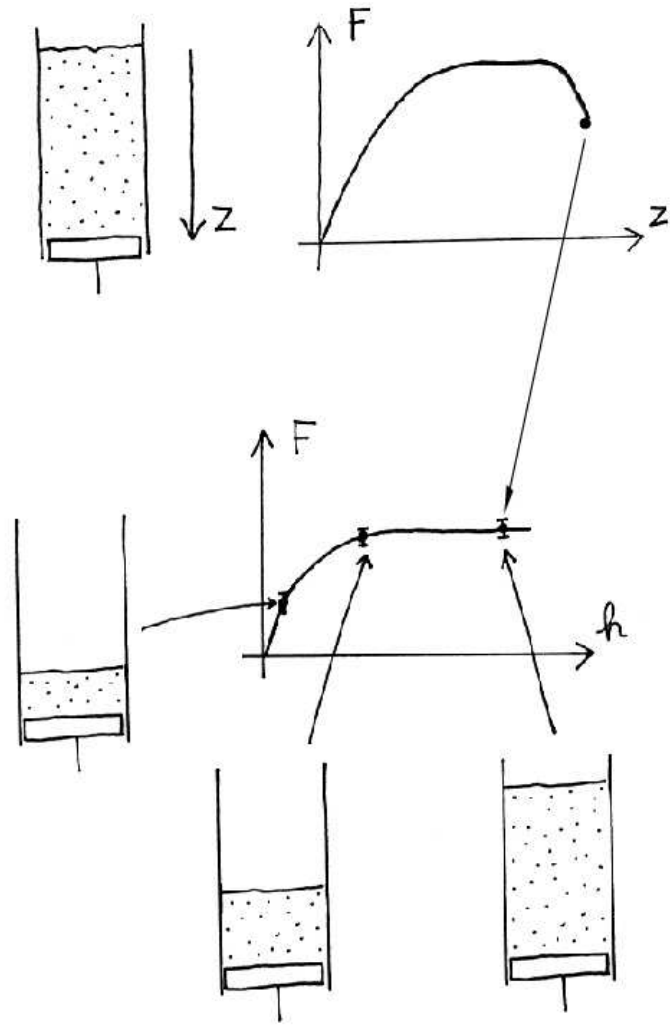}\caption{Sketch of comparison
between a pressure profile and measures at the bottom for an
elastic medium.} \label{fig7}
\end{figure}

It is thus important to note that the usual Janssen experiment, in
which one measure is made for one given height, is not equivalent
to measuring a pressure profile, and results in a lower saturation
stress (i.e. higher Janssen's constant $K$) than the pressure
profile. For more clarity on this last point, we illustrate this
difference on Fig. \ref{fig7}.

\clearpage

\begin{figure}[ptb]
\includegraphics[width=8cm]{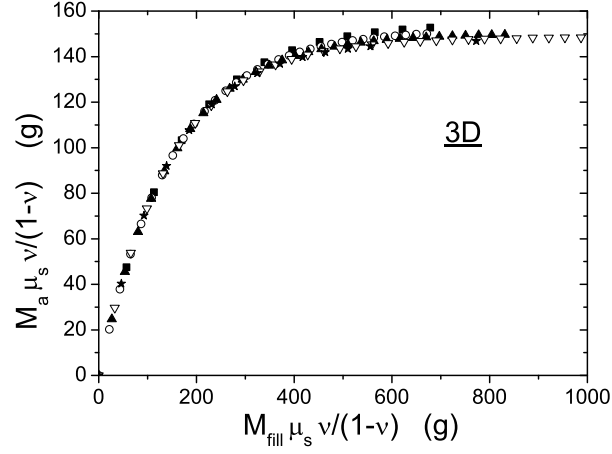}\caption{Study of the rescaling
with the Poisson ratio $\nu_{p}$ for 3D simulations of an elastic
material. We plot $M_{a}\times\mu_{s} \nu_{p}/(1-\nu_{p})$ vs.
$M_{fill}\times\mu_{s} \nu_{p}/(1\!-\!\nu_{p}) $ for
$\nu_{p}=0.26$ (squares), $\nu_{p}=0.35$ (open circles),
$\nu_{p}=0.4$ (triangles), $\nu_{p}=0.45$ (open down triangles)
and $\nu_{p}=0.49$ (stars); the friction at the walls is
$\mu_{s}=0.5$.} \label{fig8}
\end{figure}

\begin{figure}[ptb]
\includegraphics[width=8cm]{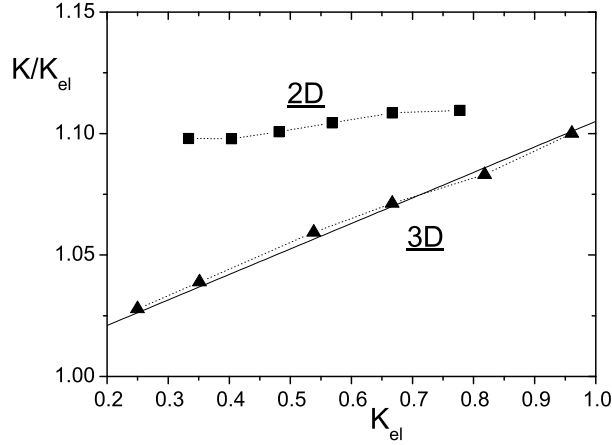}\caption{Janssen constant $K$
extracted form 2D (squares) and 3D (triangles) simulations for
various Poisson ratios. In the 2D simulations, the friction at the
walls is $\mu_{s}=1.0$; in the 3D simulations, the friction at the
walls is $\mu_{s}=0.5$. $K$ is plotted vs. the elastic stress
redirection constant $K_{el}=\nu_{p}$ in 2D,
$K_{el}=\nu_{p}/(1-\nu_{p})$ in 3D.} \label{fig9}
\end{figure}

In the following, we study the rescaling law of simulated Janssen experiments
with friction coefficient $\mu_{s}$ at the walls and the Poisson ratio
$\nu_{p}$. The data are compared to the Janssen model predictions by plotting
$M_{a}\times\mu_{s}\times\nu_{p}/(1-\nu_{p})=f(M_{fill}\times\mu_{s}\times
\nu_{p}/(1-\nu_{p}))$ for different $\mu_{s}$ and $\nu_{p}$ (since
$K_{el}\!=\!\nu_{p}/(1-\nu_{p})$ and $M_{sat}\!\propto\! 1/(K_{el}\mu_{s})$).

\subsubsection{Effect of the Poisson ratio}

On Fig. \ref{fig8}, we study the rescaling with the Poisson ratio
$\nu_{p}$ for 3D simulations. The rescaling law is rather good,
the differences may not be observable experimentally. On Fig.
\ref{fig9} we plot the Janssen coefficient $K$ extracted from the
fit of data in 2D and 3D versus the elastic stress redirection
constants $K_{el}$ ($\nu_{p}$ in 2D, $\nu_{p}/(1-\nu_{p})$ in 3D).

We observe that for a given friction coefficient, $K$ hardly depends on
$\nu_{p}$ in 2D: $K$ variation is $1\%$ for $K_{el}$ varying from $0.33$ to
$0.77$. In 3D, $K/K_{el}$ increases roughly linearly with $K_{el}$ for
$K_{el}$ varying from $0.25$ to $0.96$; however, K variation is less than 10\%
in this range. We remark that $K>K_{el}$: we explained it in the preceding
section as a consequence of the presence of the rigid bottom.

\begin{figure}[ptb]
\includegraphics[width=8cm]{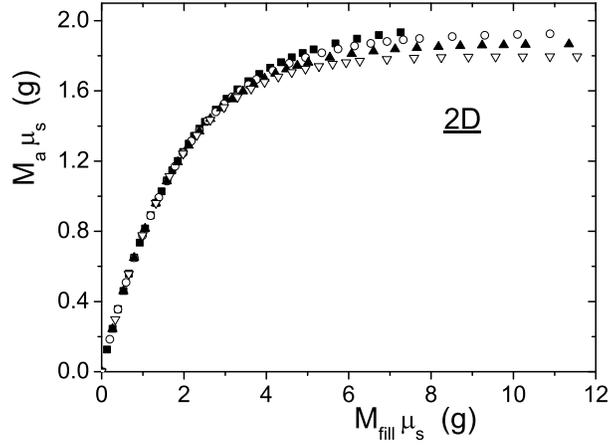}\caption{Study of the rescaling with
the friction coefficient $\mu_{s}$ at the walls for 2D simulations
of an elastic material. We plot $M_{a}\times\mu_{s}$ vs.
$M_{fill}\times\mu_{s}$ for $\mu_{s}=0.4$ (squares), $\mu_{s}=0.6$
(open circles), $\mu_{s}=0.8$ (triangles) and $\mu_{s}=1.0$ (open
down triangles); the Poisson ratio is $\nu_{p}=0.77$.}
\label{fig10}
\end{figure}

\begin{figure}[ptb]
\includegraphics[width=8cm]{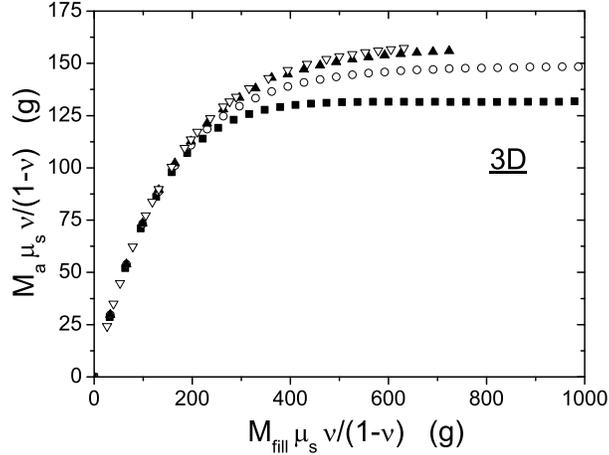}\caption{Study of the rescaling
with the friction coefficient $\mu_{s}$ at the walls for 3D
simulations of an elastic material. We plot
$M_{a}\times\mu_{s}\nu_{p}/(1-\nu_{p})$ vs.
$M_{fill}\times\mu_{s}\nu_{p}/(1-\nu_{p})$ for $\mu_{s}=0.1$ (open
down triangles) $\mu_{s}=0.25$ (triangles), $\mu_{s}=0.5$ (open
circles) and $\mu_{s}=0.8$ (squares); the Poisson ratio is
$\nu_{p}=0.45$.} \label{fig11}
\end{figure}

\subsubsection{Effect of friction at the walls}

\label{Nooverweightvaryfriction}

On Fig. \ref{fig10} and Fig. \ref{fig11}, we study the rescaling
with the friction coefficient $\mu_{s}$ at the walls respectively
for 2D and 3D simulations. We now see that the proposed rescaling
law is not good: the saturation is more abrupt as the friction
coefficient is higher.

\begin{figure}[ptb]
\includegraphics[width=8cm]{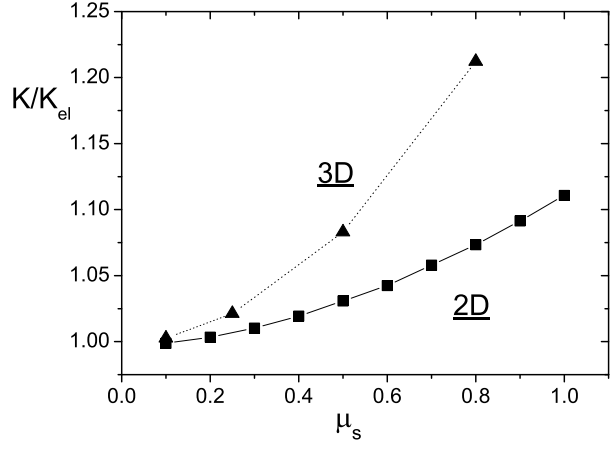}\caption{Janssen constant $K$
extracted form 2D (squares) and 3D (triangles) simulations for
various friction coefficient $\mu_{s}$. In the 2D simulations, the
Poisson ratio is $\nu_{p}=0.77$; in the 2D simulations, the
Poisson ratio is $\nu_{p}=0.45$. $K$ is rescaled by the elastic
stress redirection constant $K_{el}=0.77$ in 2D, $K_{el}=0.82$ in
3D.} \label{fig12}
\end{figure}

On Fig. \ref{fig12} we plot the Janssen coefficient $K$ extracted
from the saturation mass value in 2D and 3D versus $\mu_{s}$. We
now observe that the Janssen constant (and thus the saturation
mass) depends strongly on the friction coefficient at the walls:
$K$ increase is $12\%$ in 2D for $\mu_{s}$ varying between $0.1$
and $1.0$, and $20\%$ in 3D for $\mu_{s}$ varying from $0.1$ to
$0.8$. $K$ depends roughly quadratically on $\mu_{s}$. Moreover,
the Janssen constant seems to tend towards the elastic stress
redirection constant $K_{el}$ when $\mu_{s}$ goes to 0.

\begin{figure}[ptb]
\includegraphics[width=8cm]{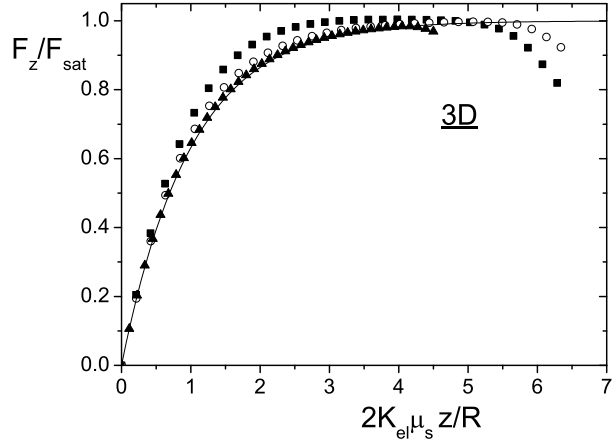}\caption{Integral of
vertical stresses $F_{z}$ vs. depth $z$, for elastic materials of
Poisson ratio $\nu_{p}=0.45$, and friction at the walls
$\mu_{s}=0.25$ (triangles), $\mu_{s}=0.5$ (open circles), et
$\mu_{s}=0.8$ (squares). We plot $F_{z}/F_{sat}$ vs.
$2K_{el}\mu_{s}\,z/R$ where $F_{sat}=\rho g\pi
R^{3}/(2K_{el}\mu_{s})$ is the theoretical saturation value for
$F_{z}$. $z/R=0$ corresponds to the top of the material. We also
plot the Janssen model prediction with
$K=K_{el}=\nu_{p}/(1-\nu_{p})$ (line).} \label{fig13}
\end{figure}

In order to understand these features, we plot on Fig. \ref{fig13}
the mean vertical pressure $F_{z}$ versus the depth, and its
rescaling with $\mu_{s}$.

We see that, regarding the pressure profile, the Janssen rescaling
law is correct for the asymptotic value, but the profiles are
slightly different: for low friction at the walls
$\mu_{s}\approx0.25$, the profile is perfectly fitted by the
Janssen law; for higher friction, the profile is sharper and
saturates abruptly. The differences for simulated Janssen
experiments observed on Fig. \ref{fig11} are now identified as a
consequence of the presence of a rigid bottom: we actually see on
Fig. \ref{fig13} that the decrease of pressure near the bottom is
more important when the friction is higher i.e. the saturation
mass is lower (and the effective Janssen constant is higher).

We now understand all these features: the parabolic part of asymptotic
displacements is negligible for low friction; in this case, the flat
displacement imposed by the rigid bottom matches the asymptotic displacement,
and the pressure at the bottom is the saturation pressure: we thus obtain
$K=K_{el}$. As the friction is increased (and the material leans on the
walls), the parabolic part of asymptotic displacement becomes more important
and the influence of the rigid bottom is to decrease pressure; we thus observe
an increase in the effective Janssen constant $K$ with $\mu_{s}$.

\subsubsection{A unique parameter}

\begin{figure}[ptb]
\includegraphics[width=8cm]{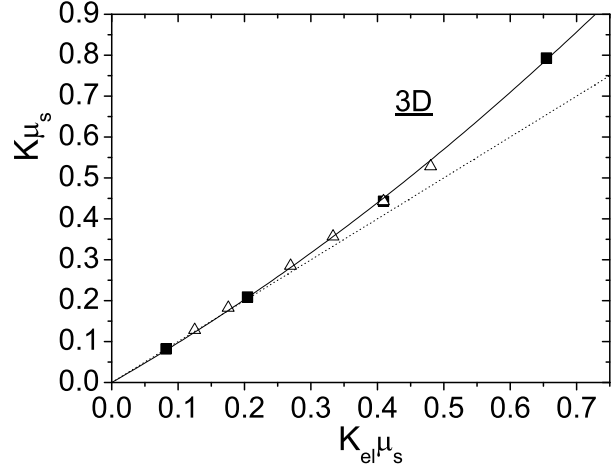}\caption{Janssen constant $K$
extracted form 3D (triangles) simulations for various friction
coefficient $\mu_{s}$ at constant $\nu_{p}=0.45$ (empty triangles)
and various Poisson ratios $\nu_{p}$ at constant $\mu_{s}=0.5$
(squares); see Fig. \ref{fig8} and \ref{fig11} for $\nu_{p}$ and
$\mu_{s}$ values. The dotted line is the $y=x$ line; the line is a
polynomial fit with $K\mu
_{s}=K_{el}\mu_{s}+0.29(K_{el}\mu_{s})^{2}$.} \label{fig14}
\end{figure}

In the Janssen picture, the Janssen coefficient $K$ and the
friction at the walls $\mu_{s}$ are not independent parameters:
the relevant parameter is $K\mu_{s}$. As regards the saturation
mass $M_{sat}$, it remains true in elasticity, and we see on Fig.
\ref{fig14} that $K\mu_{s}$, extracted from the saturation mass
value, is a function of $K_{el}\mu_{s}$ alone (i.e. data on Fig.
\ref{fig9} and \ref{fig12} can be replotted on a single universal
curve); we find a quadratic dependence of $K\mu_{s}$ on
$K_{el}\mu_{s}$: $K\mu_{s}\approx
K_{el}\mu_{s}+0.29(K_{el}\mu_{s})^{2}$. This explains the
dependence found before on $\nu_{p}$ at fixed $\mu_{s}$ or on
$\mu_{s}$ at fixed $\nu_{p}$.

\begin{figure}[ptb]
\includegraphics[width=8cm]{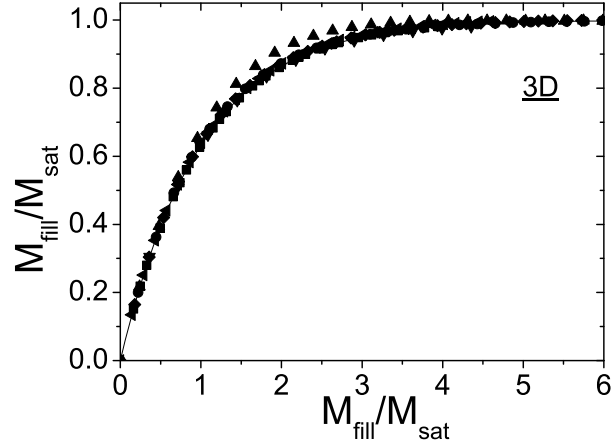}\caption{$M_{a}/M_{sat}$
vs. $M_{fill}/M_{sat}$ for 3D simulations for various values of
$\nu_{p}$ and $\mu_{s}$; the triangles are for $\nu_{p}=0.45$ and
$\mu_{s}=0.8$.} \label{fig15}
\end{figure}

However, it is not true anymore for the whole Janssen profile: we
see on Fig. \ref{fig15} that there is not a universal curve
$M_{a}/M_{sat}=f(M_{fill}/M_{sat})$: if most data may be fit by
the Janssen curve (with $f(x)=1-\exp(-x)$, it is not true anymore
for high (i.e. low $M_{sat}$ values); the data for high values of
$K\mu_{s}$ (=0.65 here) fall above the Janssen curve. However, for
most experimental conditions, $K\mu_{s}$ is not as high, and this
effect will not be observable.

\subsubsection{Effect of walls elasticity}

\begin{figure}[ptb]
\includegraphics[width=8cm]{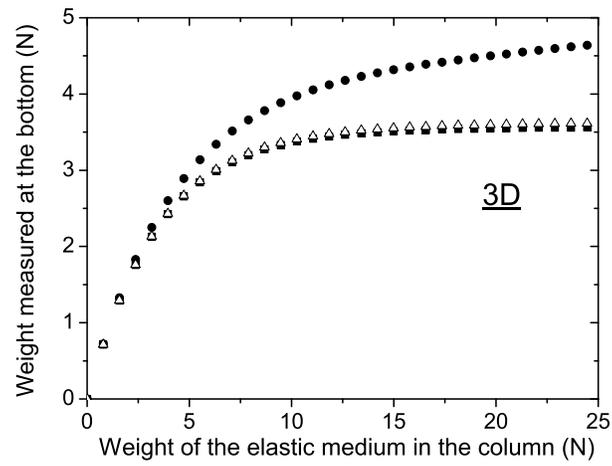}\caption{Simulation of a
Janssen experiment for elastic materials of Young modulus $E=1$
MPa (squares), $E=200$ MPa (open triangles) et $E=4$ GPa
(circles), of Poisson ratio $\nu _{p}=0.45$, in a steel column of
radius $R=4$ cm; the friction coefficient at the walls is
$\mu_{s}=0.5$. } \label{fig16}
\end{figure}

On Fig. \ref{fig16}, the Young modulus is varied.

We observe that the results are independent of the Young modulus
$E$ value, as long as it is less than 500 MPa; we see on Fig.
\ref{fig16} that the curves obtained for the simulation of a
Janssen experiment for $E=1$ MPa and $E=200$ MPa can be perfectly
superposed. For higher $E$, the weighted mass is higher and does
not seem to saturate anymore; as an example, for $E=4$ GPa on Fig.
\ref{fig16}, $M_{a}$ seems to increase indefinitely.

Note that this Young modulus effect is present only because we take into
account the walls elasticity; for rigid walls and bottom, there would not be
any dependence on $E$. The values of $E$ presented here have a meaning only
for a particular cylinder (same $E$ and same thickness).

\clearpage

\subsection{Simulations with an overweight}

In this section, we present simulations of a Janssen experiment, when an
overweight corresponding to the saturation mass is added on top of the material.

\subsubsection{Effect of the Poisson ratio}

\begin{figure}[ptb]
\includegraphics[width=8cm]{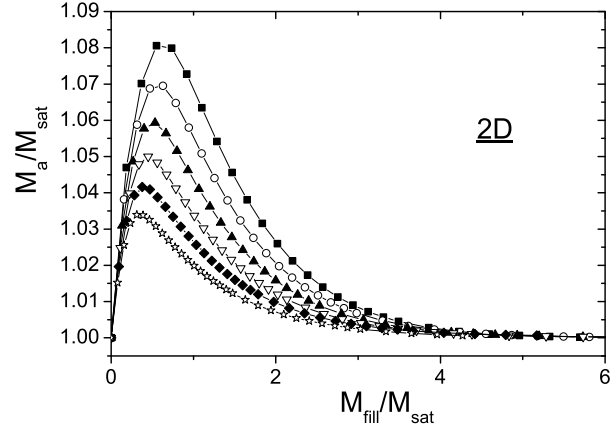}\caption{Weight at the bottom of a
2D elastic material, when an overweight equal to the saturation
mass is added on top of the material, for various Poisson ratio
$\nu_{p}$: 0.33 (open stars), 0.40 (diamonds), 0.48 (open down
triangles), 0.57 (triangles), 0.67 (open circles) and 0.78
(squares). The friction coefficient at the walls is $\mu_{s}=1.0$.
The data are scaled with the saturation mass obtained in a
simulation without overweight.} \label{fig17}
\end{figure}

\begin{figure}[ptb]
\includegraphics[width=8cm]{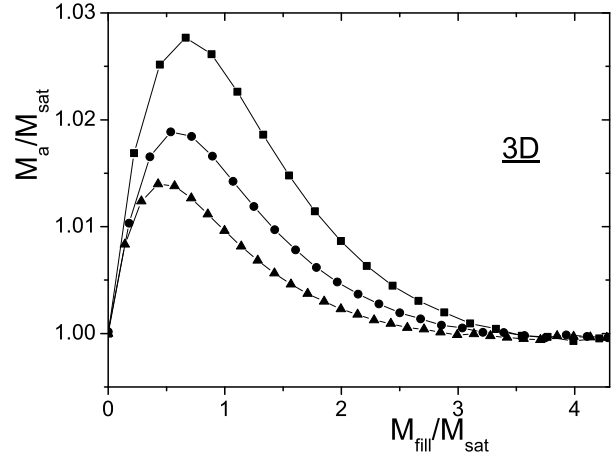}\caption{Weight at the bottom of a
3D elastic material, when an overweight equal to the saturation
mass is added on top of the material, for Poisson ratio
$\nu_{p}=0.35$ (triangles), $\nu _{p}=0.4$ (circles) et
$\nu_{p}=0.45$ (squares). The friction coefficient at the walls is
$\mu_{s}=0.5$. The data are scaled with the saturation mass
obtained in a simulation without overweight.} \label{fig18}
\end{figure}

On Fig. \ref{fig17} and \ref{fig18}, we plot the apparent mass
$M_{a}$ versus the filling mass $M_{fill}$, rescaled by the
saturation mass $M_{sat}$, respectively in 2D and 3D, for various
Poisson ratio.

We now observe that, contrary to the simulations without any overweight, the
results depend strongly on $\nu_{p}$, i.e. on the elastic stress redirection
constant $K_{el}$. The curves all have the same form: $M_{a}$ increases with
$M_{fill}$, up to a maximum $M_{max}$, then decreases slowly towards the
saturation mass $M_{sat}$. The relative maximum $M_{max}/M_{sat}$ increase
with $\nu_{p}$, and takes its value for higher $M_{fill}/M_{sat}$.

The proposed rescaling law, similar to the Janssen one, is not correct. We did
not find any rescaling law; we thus keep our rescaling as a practical
representation of data.

\subsubsection{Effect of friction at the walls}

\begin{figure}[ptb]
\includegraphics[width=8cm]{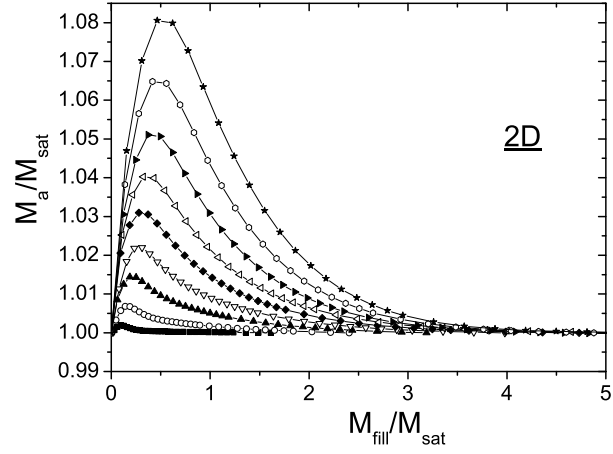}\caption{Weight at the bottom of a
2D elastic material, when an overweight equal to the saturation
mass is added on top of the material, for various friction
coefficients $\mu_{s}$: 0.2 (squares),0.3 (open circles)), 0.4
(triangles), 0.5 (open down triangles), 0.6 (diamonds), 0.7 (open
left triangles), 0.8 (right triangles), 0.9 (open hexagons) and
1.0 (stars). The Poisson ratio is $\nu_{p}=0.78$. The data are
scaled with the saturation mass obtained in a simulation without
overweight.} \label{fig19}
\end{figure}

\begin{figure}[ptb]
\includegraphics[width=8cm]{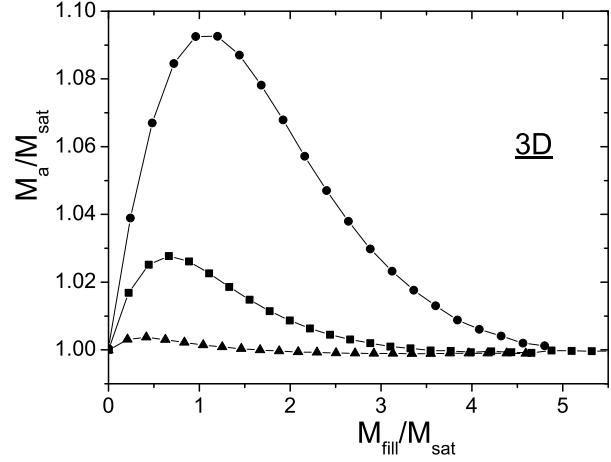}\caption{Weight at the bottom of a
3D elastic material, when an overweight equal to the saturation
mass is added on top of the material, pour friction coefficient
$\mu_{s}=0.25$ (triangles), $\mu_{s}=0.5$ (squares) and
$\mu_{s}=0.8$ (circles). The Poisson ratio is $\nu_{p}=0.45$. he
data are scaled with the saturation mass obtained in a simulation
without overweight.} \label{fig20}
\end{figure}

On Fig. \ref{fig19} et \ref{fig20}, we plot the apparent mass
$M_{a}$ versus the filling mass $M_{fill}$, rescaled by the
saturation mass $M_{sat}$, respectively in 2D and 3D, for various
friction coefficients at the walls.

We observe that the relative maximum of the apparent mass $M_{max}/M_{sat}$
increases with the friction coefficient at the walls, and takes its value for
higher $M_{fill}/M_{sat}$.

This result leads to an apparent paradox. As actually the maximum $M_{max}$
increases when $M_{sat}$ decreases (i.e. when $\mu_{s}$ and $\nu_{p}$
increase), we observe that \textit{the more the weight of the grains is
screened by the walls, the less the weight of the overload is screened}! We
will propose an interpretation in the following.

\subsubsection{Study of the pressure profile}

\begin{figure}[ptb]
\includegraphics[width=8cm]{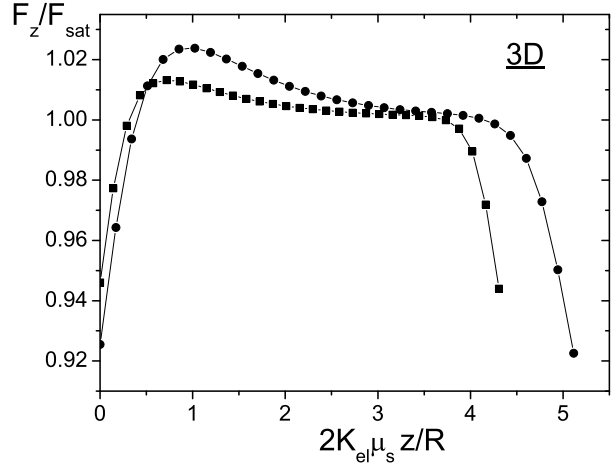}\caption{Integral of vertical
stresses $F_{z}$ vs. depth $z$, for elastic materials of friction
coefficient at the walls $\mu_{s}=0.5$, and Poisson ratios
$\nu_{p}=0.35$ (squares) and $\nu_{p}=0.45$ (circles), when an
overweight equal to the saturation mass is added on top of the
material. We plot $F_{z}/F_{sat}$ vs. $K_{el}\mu_{s}\,z/R$ where
$F_{sat}=\rho g\pi R^{3}/(2K_{el}\mu_{s})$ is the theoretical
saturation value for $F_{z}$. $z/R=0$ corresponds to the top of
the material.} \label{fig21}
\end{figure}

We can wonder if all these results remain true for the profile.
These features could be due only to the presence of a bottom. On
Fig. \ref{fig21}, we plot the mean pressure profile when an
overweight of mass equal to the saturation mass obtained in the
simulation of a Janssen experiment is added on top of the
material.

\begin{figure}[ptb]
\includegraphics[width=8cm]{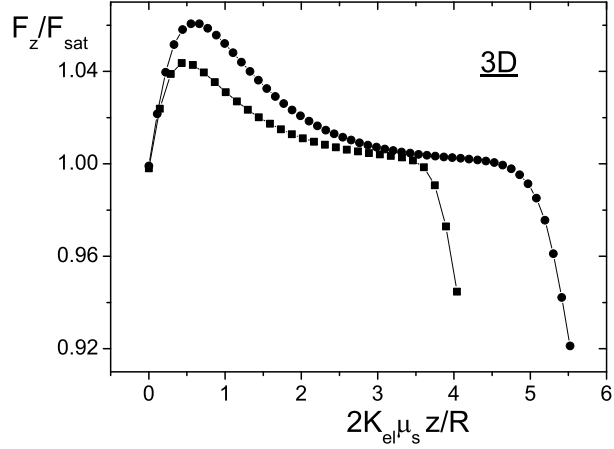}\caption{Integral of
vertical stresses $F_{z}$ vs. depth $z$, for elastic materials of
friction coefficient at the walls $\mu_{s}=0.5$, and Poisson
ratios $\nu_{p}=0.35$ (squares) and $\nu_{p}=0.45$ (circles), when
an overweight equal to the saturation value $F_{sat}$ of $F_{z}$.
We plot $F_{z}/F_{sat}$ vs. $K_{el} \mu_{s}\,z/R$ where
$F_{sat}=\rho g\pi R^{3}/(2K_{el}\mu_{s})$ is the theoretical
saturation value for $F_{z}$. $z/R=0$ corresponds to the top of
the material.} \label{fig22}
\end{figure}

We observe the same features as for the simulation of a Janssen
experiment. In order to go one step further, we now add an
overweight which imposes on top of the material a mean pressure
equal to the saturation pressure; the results are presented on
Fig. \ref{fig22}.

Once again, we observe the same features as for the simulation of a Janssen
experiment. The Janssen rescaling law is still incorrect, and the overshoot
effect is more important for higher Poisson ratio and friction coefficient.
Moreover, on the profile the amplitude of the maximum is more important: for
friction coefficient $\mu_{s}=0.5$ and Poisson ratio $\nu_{p}=0.45$, we find
that the maximum force on the profile is $1.06$ times the saturation force
$F_{sat}$, whereas the maximum weighted mass is $1.03$ times the saturation
mass $M_{sat}$.

\begin{figure}[ptb]
\includegraphics[width=8cm]{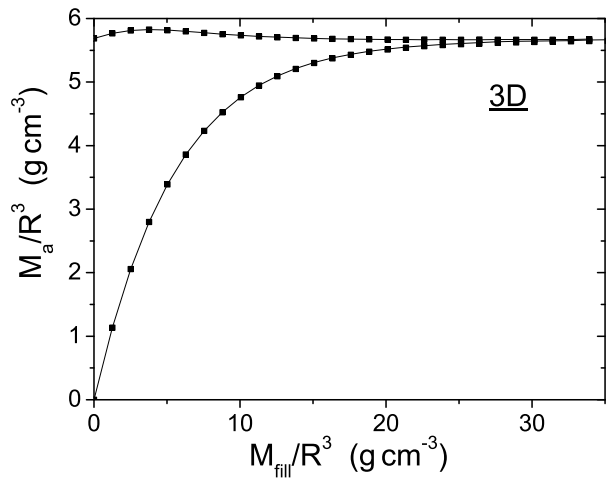}\caption{Study of the
rescaling with the radius $R$ of the column. We plot $M_{a}/R^{3}$
vs. $M_{fill}/R^{3}$ for $R=1.9$ cm (squares) et $R=4$ cm (line),
for an elastic material of Poisson ratio $\nu_{p}=0.45$ and
friction coefficient at the walls $\mu_{s}=0.5$.} \label{fig23}
\end{figure}

\subsubsection{Rescaling with the radius}

We finally verify the rescaling with the radius column for
simulations with and without any overweight. This rescaling is
perfect for both simulations (Fig. \ref{fig23}) for the extreme
radius employed in \cite{OvarlezSurp03} ($R=1.9$ cm et $R=4$ cm).

\subsection{Observable features}

To summarize, several features are experimentally observable if the isotropic
elastic theory is valid.

In a Janssen experiment with free top surface, for a same material, the
Janssen constant $K$, deduced from the measured saturation mass $M_{sat}$,
must be higher for higher friction coefficient at the walls.

In an experiment with $M_{sat}$ overweight on top of the material, the
apparent mass $M_{a}$ must increase strongly with the filling mass $M_{fill}$,
and then decrease slowly towards the saturation mass $M_{sat}$. The observed
maximum $M_{max}$ must increase with friction at the walls and with Janssen
coefficient $K$ (measured in a Janssen experiment with free top surface).

The saturation mass $M_{sat}$ deduced from the pressure profile must be higher
than the one measured at the bottom, and a strong decrease in the pressure
must be observed near the bottom. This kind of feature would not be observed
for a hyperbolic theory such as OSL \cite{Cates}. Moreover, for experiments
with an overweight on top of the material, the maximum of $M_{a}$ will be
higher on the pressure profile.

\subsection{Comparison with experimental results}

In this section, we compare the elastic theory predictions to the experimental
results obtained in \cite{OvarlezSurp03}. The main features have already been
presented in \cite{OvarlezSurp03}

\subsubsection{Classical Janssen experiment}

The Janssen experiment data are perfectly fitted by the Janssen model.
Therefore, they are also perfectly fitted by the elastic theory: we indeed
showed in Sec. \ref{Nooverweight} that for weak friction at the wall ($\mu
_{s}=0.25$ as in the experiment) the Janssen model and elastic theory
predictions cannot be discerned.\newline We also showed in Sec.
\ref{Nooverweightvaryfriction} that when the friction coefficient is
increased, the elastic theory predict a sharper initial increase of $M_{a}$
with $M_{fill}$ and an more abrupt saturation. However, this cannot be
observed experimentally for the small range of friction coefficient used in
\cite{OvarlezSurp03} (from 0.22 to 0.28). In order to test this last
prediction, it would be necessary to work with higher friction at the walls
($\mu_{s}\sim0.5$).\newline

Note that for high packing fractions ($\nu=0.645\pm0.005$), we
obtained Janssen coefficients higher than 1 ($K=1.2\pm0.1$): this
cannot be obtained in the isotropic elastic theory, as
$K_{el}=\nu_{p}/(1-\nu_{p})\leq1$. We saw that $K_{el}$ is less
than $K$ (Fig. \ref{fig9}, \ref{fig12}), due to presence of a
rigid bottom, but for small friction at the walls, this cannot
explain high $K$ values: we deduce from Fig. \ref{fig12} that for
$\mu _{s}=0.25$, $K$ must be less than 1.03; for $\mu_{s}=0.8$,
$K$ could be as high as 1.2.\newline We will see in Sec.
\ref{anisotropy} how to obtain higher $K$ values in the context of
anisotropic elasticity.

It was observed in \cite{OvarlezSurp03} that for different preparations,
characterized by different packing fractions, the saturation mass $M_{sat}$ is
lower for higher packing fraction. Although there may be structural
differences between the pilings other than the packing fraction, this can be
interpreted as increase of the Janssen coefficient $K$ with packing fraction.
Thus, in the context of isotropic elasticity, this leads to an effective
increase of the Poisson ratio $\nu_{p}$ with packing fraction. In
\cite{OvarlezSurp03}, an effective relation between packing fraction $\bar\nu$
and Poisson ratio $\nu_{p}$ was derived: $\nu_{p}\simeq2.3(\nu-0.41)$ with a
precision of $5\%$. Note that the largest packing fraction $\nu=0.645\pm0.005$
would give a Poisson ratio $\nu_{p}=0.54\pm0.03$ marginally larger than the
limit value of $1/2$, as commented above.

\subsubsection{Overweight experiment}

\begin{figure}[ptb]
\includegraphics[width=8cm]{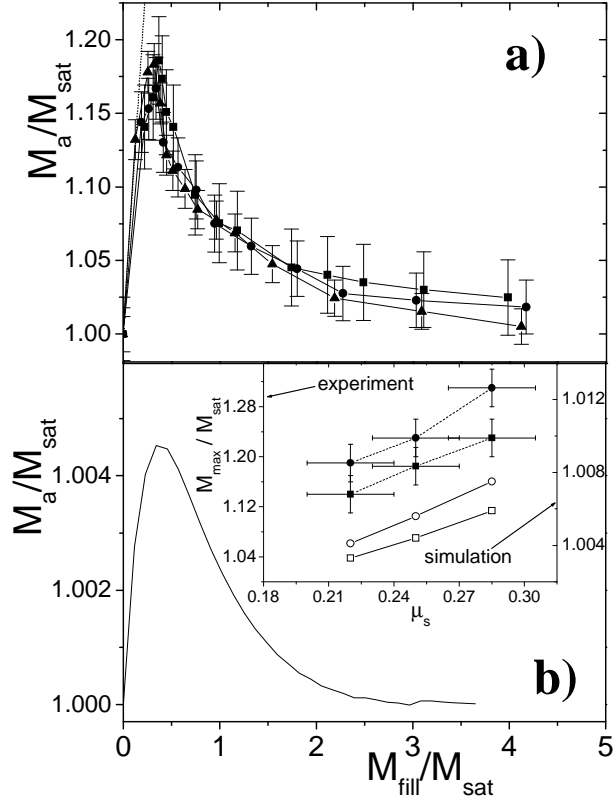}\caption{a: Apparent mass
$M_{a}$ vs. filling mass $M_{fill}$, rescaled by the saturation
mass $M_{sat} $, for loose packing in medium-rough columns of 3
diameters (38 mm (squares), 56 mm (circles), 80 mm (triangles))
with an overweight equal to $M_{sat}$; the dotted line is the
hydrostatic curve. b: Simulation of the experiment of Fig.
\ref{fig24}a, for an elastic medium characterized by the same
saturation mass (Poisson ratio $\nu_{p}=0.46$) and the same
friction at the walls ($\mu_{s}=0.25$). Inset: maximum mass
$M_{max}$ rescaled by saturation mass $M_{sat}$ vs. static
coefficient of friction at the walls $\mu_{s}$, in experiments
made on loose (squares) and dense (circles) packing in 38 mm
diameter columns, and in simulations for elastic media of Poisson
ratios $\nu_{p}=0.45$ (open squares) and $\nu_{p}=0.49$ (open
circles); the left vertical scale is used for the experimental
data, the right vertical scale is used for the simulation data.}
\label{fig24}
\end{figure}

The elastic theory gives qualitatively the same behavior as the
experimental results (see Fig. \ref{fig24}): the apparent mass
$M_{a}$ increases with the filling mass $M_{fill}$, up to a
maximum $M_{max}$, then decreases slowly towards the saturation
mass $M_{sat}$. Furthermore, all the features predicted by
elasticity can be observed: $R^{3}$ rescaling, $M_{max}/M_{sat}$
increase with friction at the walls, and $M_{max}/M_{sat}$
increase with packing fraction (experiment) or Poisson ratio
(theory). Note that $M_{max}/M_{sat}$ increases with packing
fraction and Poisson ratio are equivalent because of the effective
increase of Poisson ratio with packing fraction deduced from the
classical Janssen experiment.

However, there is no quantitative agreement between isotropic
elasticity predictions and experiments. Fig. \ref{fig24}). The
elastic curve is very similar to the experimental one, but the
experimentally observed maxima are 30 to 40 times larger.

\subsection{An apparent paradox explained by elasticity}

Interestingly, we find in the elastic case the same qualitative phenomenology
as in the experiment i.e. the computed values of the overshoot $M\max$
rescaled by the saturation mass $M_{sat}$ increases both with the friction at
the walls and the Poisson coefficient (i.e. with the effective Janssen's
parameter). This features reads as a paradox: \textit{the more the weight of
the grains is screened by the walls, the less the weight of the overload is
screened}, but we can now try to understand it at least qualitatively.

If we impose on the top surface of an elastic medium the asymptotic values for
displacements and stresses (see eq. (\ref{stressasympt}), (\ref{depasympt})),
these values then extend to the rest of the column. Thus, with such an
overweight, a flat pressure profile along depth z $\sigma_{zz}(r,z)=-\frac
{\rho gR}{2K_{el}\mu_{s}}$ is obtained as in Janssen's theory. Actually, with
the overload, the displacement imposed experimentally on the surface is almost
constant: $u_{z}(r)=u_{0}$ since the overweight is much more rigid than the
material. Then, as the asymptotic displacement is parabolic, we must have a
''transition regime'', which is at the origin of the overshoot effect.

There are two limits in which this transition can disappear, i.e. when the
Poisson coefficient $\nu$ or when friction at the walls $\mu_{s}$ are
decreased to zero. Then the parabolic part of the asymptotic displacements
(eq. (\ref{depasympt})) becomes negligible and therefore, the imposed
displacement on the surface is close to the asymptotic value. This results in
a decrease of the overshoot amplitude.

Basically, we thus recover in the elastic case, the same paradox as the
experimental situation. We now understand it as a consequence of the boundary
condition imposed experimentally by the overweight i.e. an almost constant
displacement on the surface.

It would be interesting to put overweights of different Young modulus on top
of a granular material in order to see if the overshoot amplitude decreases
when the overweight Young modulus is decreased.

\subsection{Pushing experiment}

In this section, we study the elastic theory predictions for the
force needed to push an elastic material upwards at constant
velocity in a column, and compare it to recent experimental
observations \cite{OvarlezRheo01,OvarlezRheo03} in the case of a
slowly driven granular material. The main features of this
analysis have been presented in \cite{OvarlezRheo03}.

\begin{figure}[ptb]
\includegraphics[width=8cm]{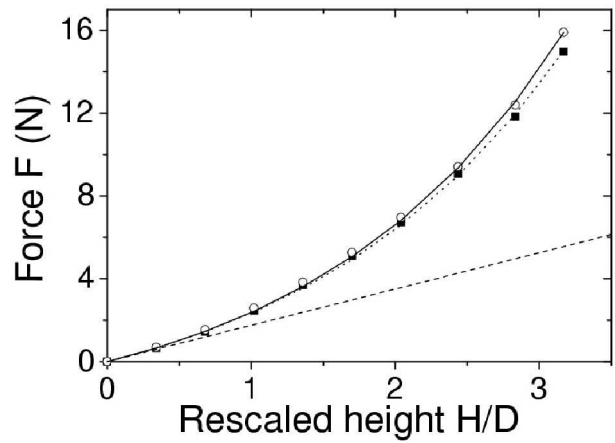}\caption{Mean resistance force
to pushing as a function of the height $H$ of the packing scaled
by the column diameter $D$, for 1.58 mm steel beads, of packing
fraction 62.5\%, in a 36 mm duralumin column, for $V=16\
\mu\mbox{m\,s}^{-1}$ (filled squares) and $V=100\
\mu\mbox{m\,s}^{-1}$ (open circles). The line and the dotted line
are fits by eq. (\ref{jansseninv}). The dashed line is the
hydrostatic curve.} \label{fig25}
\end{figure}

For a vertically pushed granular assembly
\cite{OvarlezRheo01,OvarlezRheo03} at constant velocity, the
resistance force $\bar{F}$ increases very rapidly with the
packing's height $H$(see Fig. \ref{fig25}).

Following the standard Janssen screening picture, this strong resistance to
motion is due to the leaning of the granular material on the walls (eq.
(\ref{redirection})) in association with solid friction at the side walls. At
the walls, we suppose a sliding of the granular material at a velocity $V_{0}$
(the driving velocity); the shearing stress is then
\begin{equation}
\sigma_{rz}(z)=-\mu_{d}(V_{0})\sigma_{rr}(z)
\end{equation}
where $\mu_{d}(V_{0})$ is the dynamic coefficient of friction between the
beads and the cylinder's wall at a velocity $V_{0}$.

The force $\bar{F}$ exerted by the grains on the piston can be derived from
equilibrium equations for all slices, thus we obtain:
\begin{equation}
\bar{F}=\varrho g\lambda\pi R^{2}\times(\exp(\frac{H}{\lambda})-1)
\label{jansseninv}
\end{equation}
where $\varrho$ is the mass density of the granular material, $R$ is the
cylinder radius and $g$ the acceleration of gravity. The length $\lambda
=R/2K\mu_{d}(V_{0})$ is the effective screening length.

We see on Fig. \ref{fig25} that the data are well fitted by eq.
(\ref{jansseninv}). We obtain $K\mu_{d}(V)=0.140\pm0.001$ at
$V=100\ \mu \mbox{m\,s} ^{-1}$ and $K\mu_{d}(V)=0.146\pm0.001$ at
$V=100\ \mu\mbox{m\,s} ^{-1}$.

In order to get the isotropic homogeneous elasticity prediction for the
pushing experiment, we perform a series of numerical simulations using Finite
Element Method \cite{Castem}. The condition $\sigma_{rz}=-\mu_{d}\sigma_{rr}$
is now imposed everywhere at the walls (for the pulling situation, we impose
$\sigma_{rz}=+\mu_{d} \sigma_{rr}$). The cylinder is modelled as a duralumin
elastic medium. As long as the Young modulus $E$ of the elastic medium is less
than $500$ MPa, which is usually the case for granular media, we find no
dependence of the results on $E$.

\begin{figure}[ptb]
\includegraphics[width=8cm]{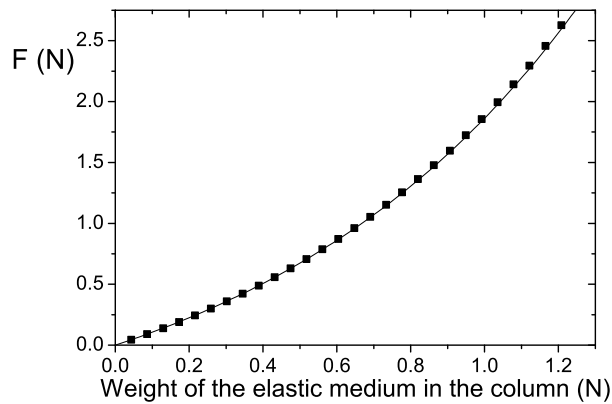}\caption{Comparison of the
resistance force to pushing simulated for a homogeneous isotropic
elastic medium (squares) of Poisson ratio $\nu=0.45$ and Young
modulus $E=100$ MPa in a duralumin cylinder of radius $R=1.9\ $cm,
with coefficient of friction $\mu_{d}=0.2$ at the walls, to the
curve obtained with eq. (\ref{jansseninv}) with Janssen
coefficient $K=K_{el}=\nu/(1-\nu)=0.82$ (line).} \label{fig26}
\end{figure}

We find no appreciable difference between the elastic prediction
(Fig. \ref{fig26}) and the curve given by eq. (\ref{jansseninv})
with $K=K_{el}$. Therefore, regarding the dependence of the
stationary state force $\bar{F}$ on the height of beads, our
system cannot be distinguished from an elastic medium.

\subsection{Pushing vs. pulling}

\begin{figure}[ptb]
\includegraphics[width=6cm,angle=-90]{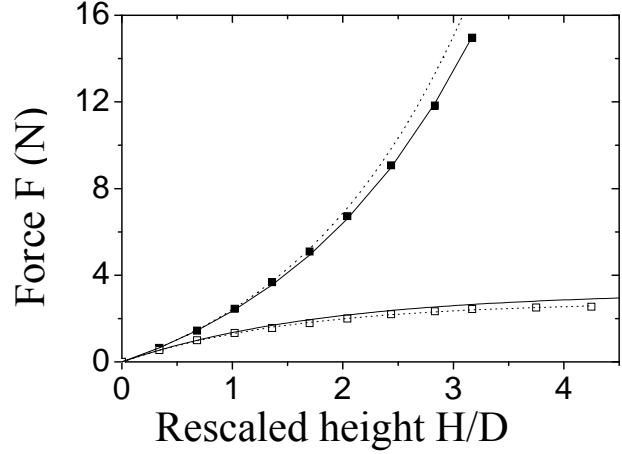}\caption{Resistance force to
pushing (filled squares) and to pulling (open squares) in the
steady-sliding regime at $V=16\ \mu\mbox{m\,s}^{-1}$ as a function
of the height $H$ of the packing scaled by the column diameter
$D$, for steel beads in the duralumin column. The lines are the
fit with eq. (\ref{jansseninv}) of the resistance force to pushing
and its prediction for the pulling situation; the dotted lines are
the fit with eq. (\ref{jansseninv}) of the resistance force to
pulling and its prediction for the pushing situation.}
\label{fig27}
\end{figure}

As a check of consistency, we performed the following dynamical
experiment in \cite{OvarlezRheo03}. First, the granular column is
pushed upwards in order to mobilize the friction forces downwards
and far enough to reach the steady state compacity. Starting from
this situation, the friction forces at the walls are reversed by
moving the piston downwards at a constant velocity $V_{down}=16\
\mu\mbox{m\,s}^{-1}$, until a stationary regime is attained. Note
that this stationary regime is characterized by the same compacity
$\overset{\_}{\nu}\approx62.5\%$ as in the pushing situation. In
Fig. \ref{fig27} the pushing force $\bar{F}$ is measured for
different packing heights $H$. The fit of experimental results
with eq. (\ref{janssen}) gives $K\mu_{d}(16\
\mu\mbox{m\,s}^{-1})=0.156\pm0.002$ which is $10\%$ larger than
$K\mu_{d}(16\ \mu\mbox{m\,s}^{-1})$ extracted from the pushing
experiment. This difference, though small, can be observed out of
uncertainties, and is systematic. It cannot be due to a slight
change in compacity $\overset{\_} {\nu}$ as from relation $\Delta
K/K\approx5\Delta\overset{\_}{\nu} /\overset{\_}{\nu}$, we would
expect a 2\% variation in compacity between the pushing and the
pulling experiment, which would be observed; we actually measured
$\Delta\overset{\_}{\nu}/\overset{\_}{\nu}=0\pm1\%$. According to
Janssen's picture, this would imply that vertical stress
redirection is more efficient in the downward pulling situation.
We believe this is a clear evidence of a granular structuring
effects but its also shows that this effect is not dominant: it
affects only $10\%$ of the average mechanical parameter $K$.

Note that finite element simulations show that the presence of a
rigid bottom implies that the effective Janssen's parameter
$K_{eff}$ extracted from Janssen's rescaling for the pulling
situation is higher than $K_{el}$, whereas for the pushing
$K_{eff}\approx K_{el}$ (as can be seen on Fig. \ref{fig26}: the
fit of the elastic curve with $K=K_{el}$ is good). Actually, if we
adjust the elastic predictions for pushing and pulling experiments
with an elastic material of Poisson coefficient $\nu_{p}=0.45$,
eq. (\ref{jansseninv}) yields a Janssen's constant $K_{eff}$ for
the pushing which is about $3\%$ lower than $K_{eff}$ for the
pulling. This is qualitatively (though not quantitatively) in
agreement with the experimental results.

Therefore isotropic elasticity can be a good framework only if we neglect the
existence of bulk restructuring effects inducing differences in the effective
Poisson coefficient of the material between the pulling and the pushing. Note
that in this case, an isotropic modelling of the granular material is somehow questionable.

\subsection{Towards anisotropy}

\label{anisotropy} We have seen in the preceding sections that isotropic
elasticity reproduces qualitatively all the features observed in the
experiments performed on granular materials. However, some problems remain: a
Janssen constant $K$ of order 1.2 was observed experimentally
\cite{OvarlezSurp03} whereas elasticity cannot provide Janssen constants
higher than 1.03 with the same experimental parameters (i.e. $\mu_{s}<0.3$);
moreover, the amplitude of the overshoot predicted by elasticity is 20 times
lower than the one observed experimentally.

That is why we study here the predictions of the simplest extension of
isotropic elasticity: transversely isotropic elasticity.

The stress-strain relations are now:

\begin{align}
\epsilon_{xx}  &
=\frac{1}{E_{1}}\;\sigma_{xx}-\frac{\nu_{1}}{E_{1}}
\;\sigma_{yy}-\frac{\nu_{2}}{E_{2}}\;\sigma_{zz}\\
\epsilon_{yy}  &
=-\frac{\nu_{1}}{E_{1}}\;\sigma_{xx}+\frac{1}{E_{1}}
\;\sigma_{yy}-\frac{\nu_{2}}{E_{2}}\;\sigma_{zz}\\
\epsilon_{zz}  &
=-\frac{\nu_{2}}{E_{2}}\;\sigma_{xx}-\frac{\nu_{2}}{E_{2}
}\;\sigma_{yy}+\frac{1}{E_{2}}\;\sigma_{zz}\\
\epsilon_{yz}  &  =\frac{1}{2G}\;\sigma_{yz}\\
\epsilon_{xz}  &  =\frac{1}{2G}\;\sigma_{xz}\\
\epsilon_{xy}  &  =\frac{(1+\nu_{1})}{E_{1}}\;\sigma_{xy}
\end{align}

Following the same calculation steps as in Sec. \ref{elastheory} , one can
show that, in the limit of high depths $z$, confinement imposes:
\begin{equation}
\sigma_{rr}=\frac{E_{1}}{E_{2}}\frac{\nu_{2}}{1-\nu_{1}}\sigma_{zz}
\end{equation}
i.e. we recover in the anisotropic case a Janssen-like relation between
stresses, with a stress redirection constant
\begin{align}
K_{anis.}=\frac{E_{1}}{E_{2}}\frac{\nu_{2}}{1-\nu_{1}}
\end{align}

The constraints on the elastic parameters are now
\begin{align}
\nu_{1}  &  >-1\\
\nu_{1}  &  <\frac{E_{1}}{E_{2}}+1\\
\nu_{2}^{2}  &  <(1-\nu_{1})\frac{E_{2}}{2E_{1}}
\end{align}

We can see the consequences on $K_{anis.}$ through an example. If $\nu_{1}=0$,
then the constraint is $\nu_{2}^{2}<\frac{E_{2}}{2E_{1}}$ which leads to
$K_{anis.}<\sqrt{\frac{E_{1}}{2E_{2}}}$. By adjusting the modulus $E_{1}$ and
$E_{2}$, one can give any value to $K_{anis.}$ which is not bounded anymore by
a maximum value of 1 (the isotropic case). This means that the experimental
values found in \cite{OvarlezSurp03} for dense packing, i.e. a Janssen
constant $K\approx1.2$, which could not be reached by the isotropic elastic
theory, can be understood in the framework of anisotropic elasticity. Note
however, that we have now 5 independent parameters instead of 2.

The problem is now that there are several ways to give $K$ a value by
adjusting independently (while satisfying the constraints) 4 parameters
$\nu_{1}$, $\nu_{2}$, $E_{1}$, and $E_{2}$; another independent parameter is
$G$, which may affect the Janssen profile shape; thus one has to find a
physical justification for choosing these values.

We tried to vary independently most parameters while keeping $K$ and $\mu_{s}$
constant in order to see if for a given $M_{sat}$, there is a way to obtain
the giant overshoot we observe experimentally.

\begin{figure}[ptb]
\includegraphics[width=8cm]{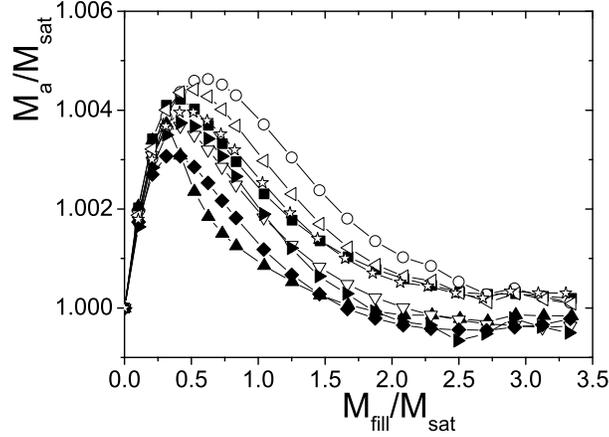}\caption{Apparent mass $M_{a}$ vs.
filling mass $M_{fill}$, rescaled by the saturation mass
$M_{sat}$, for various transversely isotropic elastic media
characterized by the same saturation mass $M_{sat}$. Squares:
$G=0.345$ MPa, $E_{1}=E_{2}=1$ MPa, $\nu_{1}=\nu_{2}=0.45$
(isotropic medium). Open circles: $G=0.1$ MPa, $E_{1}=E_{2}=1$
MPa, $\nu_{1}=\nu_{2}=0.45$. Triangles: $G=1$ MPa, $E_{1}=E_{2}=1$
MPa, $\nu_{1}=\nu_{2}=0.45$. Open down triangles: $G=0.345$ MPa,
$E_{1}=2$ MPa, $E_{2}=1$ MPa, $\nu_{1}=\nu_{2}=0.29$. Diamonds:
$G=0.345$ MPa, $E_{1}=2$ MPa, $E_{2}=1$ MPa, $\nu_{1}=-0.34$,
$\nu_{2}=0.55$. Open left triangles: $G=0.345$ MPa, $E_{1}=2$ MPa,
$E_{2}=1$ MPa, $\nu_{1}=0.756$, $\nu_{2}=0.1$. Right triangles:
$G=0.345$ MPa, $E_{1}=1$ MPa, $E_{2}=2$ MPa, $\nu_{1}=0.633$,
$\nu_{2}=0.6$. Open stars: $G=0.345$ MPa, $E_{1}=1$ MPa, $E_{2}=2$
MPa, $\nu_{1}=0.815$, $\nu_{2}=0.3$. In all these simulations, the
friction at the walls is $\mu_{s}=0.25$.} \label{fig28}
\end{figure}

Several tries are compared with the isotropic case on Fig.
\ref{fig28}. As far as we could see, there is only a rather small
influence of anisotropy on the overshoot amplitude: the deviation
from the isotropic value is at the most 15\% for reasonable values
of the modulus. Nevertheless, there are some general tendencies:
the higher the shear modulus $G$, the smaller the overshoot
amplitude is; the higher $\nu_{1}$, the higher the amplitude is,
whatever the stiffer direction may be. However, the effect is far
from sufficient to reproduce experimental results.

\subsection{A stress induced anisotropy?}

\label{anisotropy}

In order to account for the height of the experimentally observed
overshoot, we propose a toy model based on the idea of
stress-induced anisotropy. It is possible that the overweight
induces locally a change in the structure. We would expect an
increase of the number of contacts i.e. of the young modulus
$E_{2}$ in the vertical direction, if the granular material can be
modelled as an effective elastic medium. From the relation
$K_{anis.}=\frac{E_{1}}{E_{2} }\frac{\nu_{2}}{1-\nu_{1}}$, we then
expect the Janssen coefficient to be lower near the overweight
than in the bulk.

Let the Janssen constant be $K_{1}$ near the overweight, down to a
depth $H_{1}$, and $K_{2}>K_{1}$ for any depth $z>H_{1}$. The
saturation mass, which is the overweight mass, was measured with
an homogeneous material of Janssen constant $K_{2}$ and is thus
$M_{sat_{2}}=\frac{\rho\pi R^{3}}{2K_{2}\mu_{s}} $. Then, it
underestimates the saturation mass $M_{sat_{1}}=\frac{\rho\pi
R^{3}}{2K_{1}\mu_{s}}$ of the material layer near the overweight:
the measured mass $M_{a}$ then first increases with the filling
mass $M_{fill}$ and would naturally tend to a higher value
$M_{sat_{1}}$. But for a depth $H_{1}$ the material's structure is
no more affected by the overweight and is now characterized by a
Janssen constant $K_{2}$: the mass imposed on the material 2 by
the material 1 is then higher than its saturation mass
$M_{sat_{2}}$, and the measured mass $M_{a}$ has to decrease with
the filling mass $M_{fill}$ in order to reach its saturation value
$M_{sat_{2}}$.

This simple idea can be easily formalized for an ideal Janssen material. The
equilibrium equation on the vertical stress $\sigma_{zz}(z)$ in material $i$
characterized by Janssen constant $K_{i}$ reads
\begin{align}
\frac{\partial\sigma_{zz}}{\partial z}+\frac{2K_{i}\mu_{s}}{R}\sigma
_{zz}=-\rho g
\end{align}

In the material 1, for $0<z<H_{1}$, we solve this equation for the mass
$M(z)=\sigma_{zz}(z)/(\pi R^{2}g)$ weighted at depth $z$, with the boundary
condition $M(z\!=\!0)=M_{sat_{2}}$ and get
\begin{align}
M(z)=M_{sat_{1}}\biggl(1-\exp\Bigl(-2K_{1}\mu_{s}\frac{z}{R}\Bigr)\biggr
)+M_{sat_{2}}\exp\Bigl(-2K_{1}\mu_{s}\frac{z}{R}\Bigr)
\end{align}

In the material 2, for $H_{1}<z$ we solve this equation with the
boundary condition
$M(z\!=\!H_{1})=M_{sat_{1}}\bigl(1-\exp(-2K_{1}\mu_{s}\frac{H_{1}
}{R})\bigr)+M_{sat_{2}}\exp(-2K_{1}\mu_{s}\frac{H_{1}}{R})$ and
get
\begin{align}
M(z)=M_{sat_{2}}+(M_{sat_{1}}-M_{sat_{2}})\biggl(1-\exp\Bigl(-2K_{1}\mu
_{s}\frac{H_{1}}{R}\Bigr)\biggr)\exp\Bigl(-2K_{2}\mu_{s}\frac{(z-H_{1})}
{R}\Bigr)
\end{align}

These equations can be rewritten using $M_{a}/M_{sat_{2}}$ and
$M_{fill} /M_{sat_{2}}$ as variables:
\begin{align}
0<z<H_{1}  &
\rightarrow\frac{M_{a}}{M_{sat_{2}}}=\frac{M_{sat_{1}}
}{M_{sat_{2}}}\biggl(1-\exp\Bigl(-\frac{M_{fill}}{M_{sat_{2}}}\frac
{M_{sat_{2}}}{M_{sat_{1}}}\Bigr)\biggr)+\exp\Bigl(-\frac{M_{fill}}{M_{sat_{2}
}}\frac{M_{sat_{2}}}{M_{sat_{1}}}\Bigr)\\
H_{1}<z  &
\rightarrow\frac{M_{a}}{M_{sat_{2}}}=1+\bigl(\frac{M_{sat_{1}}
}{M_{sat_{2}}}-1\bigr)\biggl(1-\exp\Bigl(-\frac{H_{1}}{\lambda_{2}}
\frac{M_{sat_{2}}}{M_{sat_{1}}}\Bigr)\biggr)\exp\Bigl(-\bigl(\frac{M_{fill}
}{M_{sat_{2}}}-\frac{H_{1}}{\lambda_{2}}\bigr)\Bigr)
\end{align}
The two independent variables one can adjust are
$\frac{M_{sat_{1}} }{M_{sat_{2}}}$ (which is the Janssen constant
ratio $K_{2}/K_{1}$), and the ratio
$H_{1}/\lambda_{2}=M_{fill}(H_{1})/M_{sat_{2}}$ where $\lambda_{2}
=\rho\pi R/(2K_{2}\mu_{s})$ is the Janssen screening length in the
medium 2.

\begin{figure}[ptb]
\includegraphics[width=8cm]{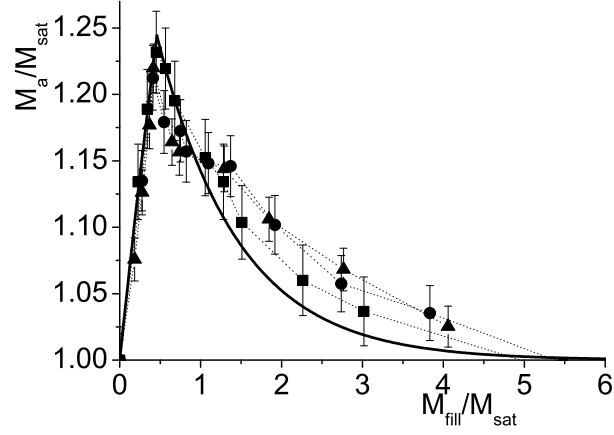}\caption{Apparent mass $M_{a}$ vs.
filling mass $M_{fill}$, rescaled by the saturation mass
$M_{sat}$, for dense packing in medium-rough columns of 3
diameters (38 mm (squares), 56 mm (circles), 80 mm (triangles))
with an overweight equal to $M_{sat}$. The data are compared with
our inhomogeneous Janssen model with $K_{1}/K_{2}=2.4$, and
$H_{1}/\lambda_{2}=0.45$.} \label{fig29}
\end{figure}

\begin{figure}[ptb]
\includegraphics[width=8cm]{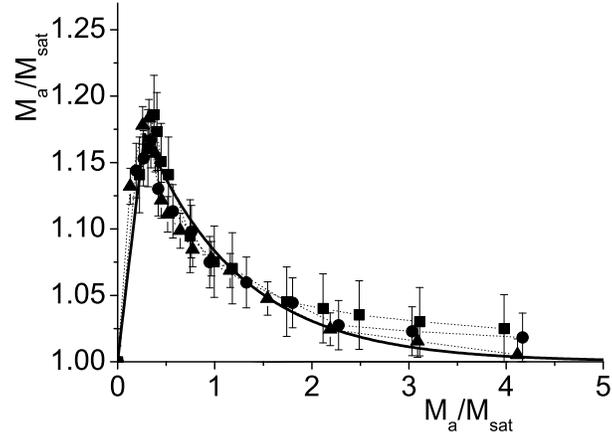}\caption{Apparent mass $M_{a}$ vs.
filling mass $M_{fill}$, rescaled by the saturation mass
$M_{sat}$, for loose packing in medium-rough columns of 3
diameters (38 mm (squares), 56 mm (circles), 80 mm (triangles))
with an overweight equal to $M_{sat}$. The data are compared with
our inhomogeneous Janssen model with $K_{1}/K_{2}=2.8$, and
$H_{1}/\lambda_{2}=0.3$. } \label{fig30}
\end{figure}

On a $M_{a}/M_{sat_{2}}=f(M_{fill}/M_{sat_{2}})$ plot, the
parameter $H_{1}/\lambda_{2}$ is the X-axis value
$M_{fill}/M_{sat_{2}}$ at which the weighted mass starts to
decrease. The slope at the origin is $1-\frac
{M_{sat_{2}}}{M_{sat_{1}}}$. Typically (see Fig. \ref{fig29} and
Fig. \ref{fig30}), the experimental slopes are of order 0.5 (it
would be 1 for an hydrostatic pressure), i.e. the Janssen
coefficient $K_{1}$ has to be about 2 times $K_{2}$; in the
anisotropic elasticity framework, it would mean that the young
modulus in the vertical direction is doubled by the overweight.
The order of the extension $H_{1}/\lambda_{2}$ of the induced
anisotropy is of order 0.4. The experimental data for dense and
loose packing are compared on Fig. \ref{fig29} and Fig.
\ref{fig30} with the model. It appears to reproduce correctly the
data; however, the parameters cannot be determined accurately as
they are obtained from the very beginning of the data (the slope
at the origin and the X-axis value of the maximum) which cannot be
measured with a high precision. Therefore, the parameters
presented in these figures are just given as examples.

\section{Conclusion}

In conclusion, we performed an extensive study of the Janssen's
column problem. It is a mixture of numerical studies both in 2D
and in 3D, in the case of frictional boundaries. The aim was to
test thoroughly the classical and celebrated Janssen's analysis in
the context of an effective homogeneous elastic material and
provide some meaning to the effective Janssen's constant of stress
redirection at it can be obtained experimentally. Note that here
we do not make any assumption on a plastic threshold in the bulk
as it is usually considered to provide bounds on the Janssen's
constant (active and passive limits). Also this analysis was
performed in the context of an extensive experimental work where
preparation was varied and were special care about friction at the
wall was taken to be able to establish a precise comparison with
theoretical modelling. Interestingly, we find that the Janssen's
approach is fully valid in the limit of low friction coefficients
between the grains and the wall (up to a moderate value of 0.5).
It means that an exponential saturation curve of the average
normal stress at the bottom is an excellent approximation, which
defines precisely an effective Janssen's constant solely dependent
on the Poisson ratio. We also derive ion the limit of an infinite
column a value for an elastic Janssen's constant ($K_{el}=\nu$ in
2D and $K_{el}=\nu/(1-\nu)$ in 3D). For a finite size column, the
presence of a bottom diminish the average vertical stress such as
to yield an effective Janssen's constant if the bottom normal
stress is measured with a value directly related to the elastic
constant $K_{el}$. Therefore, in this context experiment data can
be matched with isotropic elasticity if the packing fraction
representing the preparation can be associated with an effective
Poisson ratio.

Consequently, there is a need to provide a more strained test to
the isotropic elastic theory. This was done experimentally by
imposing on the top of the column an overweight equal to the
saturation stress \cite{OvarlezSurp03} and here we propose the
same test in the same conditions for an elastic material. The
numerical simulations show that stresses at the bottom also
exhibits an overshoot when the column depth is increased. The
relative value of the overshoot is at most 7\% . We propose a
scaling relation of the overshoot amplitude when the wall fiction
is varied. This result contrasts with a Janssen's analysis where a
flat profile should be observed. This overshoot effect is related
to the deformation of the elastic medium below the overweight
which sets a length where the deformation profile is not parabolic
any more as in the rest (more like a flat profile as it is imposed
by the overweight boundary condition). But at the quantitative
level, the overshoot effect found experimentally has an amplitude
about 20 to 30 times larger! This lack of quantitative agreement
opens new questions on the modification on the influence of the
medium due to the overload. This is the reason why we push further
the investigation in the context of anisotropic elasticity. We
consider an orthotropic elastic medium with the main stiff
direction along the vertical. In this case the effective Janssen's
redirection coefficient can be changed according to the stiffness
ratio but the overshoot test does not produce an overshoot value
significantly larger than the isotropic situation.

Finally, we propose a qualitative model based on an extension of
the Janssen's approach where we assume that the overload has
changed the medium within a given depth such as to yield a smaller
Janssen's constant. This would be consistent with the onset of
strain induced anisotropy producing a stiffer medium in the
vertical direction. The agreement of this simple model (with two
fitting parameters) is satisfactory but more importantly it raises
interesting questions and calls for new experimental work. In this
frame of mind it would be very interesting to see differences of
the stress saturation curve for two media prepared i) with a
regular rain like pouring as before and ii) a rain like deposition
process but where an overweight is imposed above each deposition
step (a deposition step being of a height much smaller than the
final height). In this case the stress induced texture changes
could provide a rational explanation to the extension of the
Janssen's model that fits correctly the overweight experiments.

We thank Profs. R.P. Behringer and J. Socolar for many fruitful
discussions.


\begin{thebibliography}{99}
\bibitem{PDM}\textit{Physics of Dry Granular Media}, ed. by H.J. Herrmann,
J.-P. Hovi and S. Luding, Kluwer Acad. Publisher (1998).

\bibitem {Oda}M.Oda, Mechanics of Materials \textbf{16}, 35 (1993).

\bibitem {Radjai}F. Radjai, D.E. Wolf, M. Jean and J.J. Moreau, Phys. Rev.
Lett. \textbf{80}, 61 (1998).

\bibitem {Geng03}J. Geng, G. Reydellet, E. Cl\'{e}ment and R.P. Behringer,
Physica D \textbf{182}, 274 (2003).

\bibitem {Attman04a}APF Attman, P. Brunet, J. Geng, G. Reydellet, G. Combe,
P. Claudin, R.P. Behringer, E. Cl\'{e}ment, \textit{to be
published in} J.Cond Mat A (2005); Preprint Cond-mat/0411734.

\bibitem {Nagel}A.J. Liu, S.R. Nagel, Nature 396, 21 (1998).

\bibitem {Cates}M.E. Cates, J.P. Wittmer, J.-P. Bouchaud and P. Claudin, Phys.
Rev. Lett. \textbf{81}, 1841 (1998).

\bibitem {Wood}D.M. Wood, \emph{Soil Behaviour and Critical State Soil
Mechanics} (Cambridge University, Cambridge, England, 1990).

\bibitem {Vaneltroudutas}L. Vanel et al., Phys. Rev. E \textbf{60}, R5040 (1999).

\bibitem {OSL}J.P. Wittmer, M.E. Cates, P. Claudin, J. Phys. I 7, 39 (1997);
J.P. Wittmer et al., Nature \textbf{382} , 336 (1996).

\bibitem {Guiguir}G. Reydellet, E. Cl\'{e}ment, Phys. Rev. Lett. \textbf{86},
3308 (2001); J. Geng et al., Phys. Rev. Lett. \textbf{87}, 035506 (2001); D.
Serero et al., Eur. Phys. J. E, \textbf{6}, 169 (2001).

\bibitem {Janssen}H.A. Jannsen, Zeitschr. D. Vereines Deutscher Inginieure
\textbf{39}, 1045 (1895).

\bibitem {Vaneljanssen}L. Vanel, E. Cl\'{e}ment, Eur. Phys. J. B \textbf{11},
525 (1999). L. Vanel et al., Phys. Rev. Lett. \textbf{84}, 1439
(2000).

\bibitem {OvarlezRheo01}G. Ovarlez, E. Kolb, E. Cl\'{e}ment, Phys. Rev. E
\textbf{64}, 060302(R) (2001).

\bibitem {OvarlezRheo03}G. Ovarlez, E. Cl\'{e}ment, Phys. Rev. E \textbf{68},
031302 (2003).

\bibitem {OvarlezSurp03}G. Ovarlez, C. Fond, E. Cl\'{e}ment, Phys. Rev. E
\textbf{67}, 060302(R) (2003).

\bibitem {Kolb99}E. Kolb, T. Mazozi, E. Clement, and J. Duran, Eur. Phys. J.
B, \textbf{8}, 483 (1999).

\bibitem {bertho02}Y. Bertho, F. Giorgiutti-Dauphine, and J.-P. Hulin, Phys. Rev.Lett. \textbf{90}, 144301 (2002).

\bibitem {Zenith03}D. Arroyo-Cetto, G. Pulos, R. Zenit and M. A.
Jimenez-Zapata, C. R. Wassgren, Phys. Rev. E \textbf{68}, 051301
(2003)

\bibitem {landry04} J.W. Landry and G.S. Grest, Phys. Rev. E \textbf{69},
031303, (2004).

\bibitem {Radjai04} I. Bratberg, F. Radjai, and A. Hansen, \textit{to be
published in} Phys. Rev. E.

\bibitem {Allain} D.Senis, C.Allain, Phys. Rev. E \textbf{55}, 7797 (1997).

\bibitem {Evesque}P. Evesque, P.-G. de Gennes, C. R. Acad. Sci. Paris,
\textbf{326} IIb, 761 (1998).

\bibitem {Castem}with CAST3M, http://www-cast3m.cea.fr

\bibitem {Elastochaines1}C. Goldenberg, I. Goldhirsch, Phys. Rev. Lett. 89,
084302 (2002).

\end{thebibliography}
\end{document}